\documentclass[twocolumn]{aastex631}

\newcommand{\apg}{\ga}
\newcommand{\apll}{\la}

\newcommand{\no}{\nodata}
\newcommand{\Msol}{\hbox{M$_\odot$}}
\newcommand{\zsp}{\ensuremath{z_{\rm sp}}}
\newcommand{\zph}{\ensuremath{z_{\rm ph}}}
\newcommand{\tn}[1]{\rlap{\tablenotemark{#1}}}
\newcommand{\EW}{\rm EW}

\shorttitle{JWST observations of G165}
\shortauthors{Frye, Pascale, Pierel et al.}

\begin{document}

\title{The JWST Discovery of the Triply-imaged Type Ia ``Supernova H0pe" and Observations of the Galaxy Cluster PLCK G165.7+67.0}

\author[0000-0003-1625-8009]{Brenda L.~Frye}
\affiliation{Department of Astronomy/Steward Observatory, University of Arizona, 933 N. Cherry Avenue, Tucson, AZ 85721, USA}

\author[0000-0002-2282-8795]{Massimo Pascale}
\affiliation{Department of Astronomy, University of California, 501 Campbell Hall \#3411, Berkeley, CA 94720, USA}

\author{Justin Pierel}
\affiliation{Space Telescope Science Institute, 3700 San Martin Drive, Baltimore, MD 21218, USA}

\author{Wenlei Chen}
\affiliation{School of Physics and Astronomy, University of Minnesota, Minneapolis, MN 55455, USA}

\author{Nicholas Foo}
\affiliation{School of Earth and Space Exploration, Arizona State University, Tempe, AZ 85287-1404, USA}

\author[0009-0001-7446-2350]{Reagen Leimbach}
\affiliation{Department of Astronomy/Steward Observatory, University of Arizona, 933 N. Cherry Avenue, Tucson, AZ 85721, USA}

\author[0000-0003-3418-2482]{Nikhil Garuda}
\affiliation{Department of Astronomy/Steward Observatory, University of Arizona, 933 N. Cherry Avenue, Tucson, AZ 85721, USA}

\author[0000-0003-3329-1337]{Seth H. Cohen} 
\affiliation{School of Earth and Space Exploration, Arizona State University,
Tempe, AZ 85287-1404, USA}

\author[0000-0001-9394-6732]{Patrick S. Kamieneski}
\affiliation{School of Earth and Space Exploration, Arizona State University,
Tempe, AZ 85287-1404, USA}

\author[0000-0001-8156-6281]{Rogier A. Windhorst} 
\affiliation{School of Earth and Space Exploration, Arizona State University, Tempe, AZ 85287-1404, USA}

\author[0000-0002-6610-2048]{Anton M. Koekemoer} 
\affiliation{Space Telescope Science Institute,
3700 San Martin Drive, Baltimore, MD 21218, USA}
    `
\author{Pat Kelly}
\affiliation{Minnesota Institute for Astrophysics, University of Minnesota, 116 Church St SE, Minneapolis, MN 55455, USA}

\author[0000-0002-7265-7920]{Jake Summers} 
\affiliation{School of Earth and Space Exploration, Arizona State University,
Tempe, AZ 85287-1404, USA}

\author{Michael Engesser}
\affiliation{Space Telescope Science Institute, 3700 San Martin Drive, Baltimore, MD 21218, USA}

\author[0000-0001-9773-7479]{Daizhong Liu}
\affiliation{Max-Planck-Institut f\"ur Extraterrestrische Physik (MPE), Giessenbachstr. 1, D-85748 Garching, Germany}

\author[0000-0001-6278-032X]{Lukas J. Furtak}
\affiliation{Physics Department, Ben-Gurion University of the Negev, P. O. Box 653, Be’er-Sheva, 8410501, Israel}

\author[0000-0001-7411-5386]{Maria del Carmen Polletta}
\affiliation{INAF – Istituto di Astrofisica Spaziale e Fisica cosmica (IASF) Milano, Via A. Corti 12, 20133 Milan, Italy}

\author[0000-0001-6342-9662]{Kevin C. Harrington} 
\affiliation{European Southern Observatory, Alonso de C\'ordova 3107, Vitacura, Casilla 19001, Santiago de Chile, Chile} 

\author[0000-0002-9895-5758]{S. P. Willner}
\affiliation{Center for Astrophysics \textbar\ Harvard \& Smithsonian, 60 Garden Street, Cambridge, MA 02138, USA}

\author[0000-0001-9065-3926]{Jose M. Diego}
\affiliation{Instituto de Fisica de Cantabria (CSIC-C). Avda. Los Castros s/n. 39005 Santander, Spain}

\author[0000-0003-1268-5230]{Rolf A. Jansen} 
\affiliation{School of Earth and Space Exploration, Arizona State University, Tempe, AZ 85287-1404, USA}

\author[0000-0001-7410-7669]{Dan Coe} 
\affiliation{Space Telescope Science Institute, 3700 San Martin Drive, Baltimore, MD 21218, USA}
\affiliation{Association of Universities for Research in Astronomy (AURA) for the European Space Agency (ESA), STScI, Baltimore, MD 21218, USA}
\affiliation{Center for Astrophysical Sciences, Department of Physics and Astronomy, The Johns Hopkins University, 3400 N Charles St. Baltimore, MD 21218, USA}

\author[0000-0003-1949-7638]{Christopher J. Conselice} 
\affiliation{Jodrell Bank Centre for Astrophysics, Alan Turing Building,
University of Manchester, Oxford Road, Manchester M13 9PL, UK}

\author[0000-0003-2091-8946]{Liang Dai}
\affiliation{Department of Physics, University of California, 366 Physics North MC 7300, Berkeley, CA. 94720, USA}

\author{Herv\'{e} Dole}
\affiliation{Universit\'e Paris-Saclay, CNRS, Institut d'Astrophysique Spatiale, 91405, Orsay, France}

\author[0000-0002-9816-1931]{Jordan C. J. D'Silva} 
\affiliation{International Centre for Radio Astronomy Research (ICRAR) and the
International Space Centre (ISC), The University of Western Australia, M468,
35 Stirling Highway, Crawley, WA 6009, Australia}
\affiliation{ARC Centre of Excellence for All Sky Astrophysics in 3 Dimensions
(ASTRO 3D), Australia}

\author[0000-0001-9491-7327]{Simon P. Driver} 
\affiliation{International Centre for Radio Astronomy Research (ICRAR) and the
International Space Centre (ISC), The University of Western Australia, M468,
35 Stirling Highway, Crawley, WA 6009, Australia}

\author[0000-0001-9440-8872]{Norman A. Grogin} 
\affiliation{Space Telescope Science Institute,
3700 San Martin Drive, Baltimore, MD 21218, USA}

\author[0000-0001-6434-7845]{Madeline A. Marshall} 
\affiliation{National Research Council of Canada, Herzberg Astronomy \&
Astrophysics Research Centre, 5071 West Saanich Road, Victoria, BC V9E 2E7,
Canada}
\affiliation{ARC Centre of Excellence for All Sky Astrophysics in 3 Dimensions
(ASTRO 3D), Australia}

\author[0000-0002-7876-4321]{Ashish K. Meena}
\affiliation{Physics Department, Ben-Gurion University of the Negev, P. O. Box 653, Be’er-Sheva, 8410501, Israel}

\author[0000-0001-6342-9662]{Mario Nonino} 
\affiliation{INAF-Osservatorio Astronomico di Trieste, Via Bazzoni 2, 34124
Trieste, Italy} 

\author[0000-0002-6150-833X]{Rafael {Ortiz~III}} 
\affiliation{School of Earth and Space Exploration, Arizona State University,
Tempe, AZ 85287-1404, USA}

\author[0000-0003-3382-5941]{Nor Pirzkal} 
\affiliation{Space Telescope Science Institute,
3700 San Martin Drive, Baltimore, MD 21218, USA}

\author[0000-0003-0429-3579]{Aaron Robotham} 
\affiliation{International Centre for Radio Astronomy Research (ICRAR) and the
International Space Centre (ISC), The University of Western Australia, M468,
35 Stirling Highway, Crawley, WA 6009, Australia}

\author[0000-0003-0894-1588]{Russell E. Ryan, Jr.} 
\affiliation{Space Telescope Science Institute,
3700 San Martin Drive, Baltimore, MD 21218, USA}

\author{Lou Strolger}
\affiliation{Space Telescope Science Institute,
3700 San Martin Drive, Baltimore, MD 21218, USA}

\author[0000-0001-9052-9837]{Scott Tompkins} 
\affiliation{School of Earth and Space Exploration, Arizona State University,
Tempe, AZ 85287-1404, USA}

\author[0000-0001-9262-9997]{Christopher N. A. Willmer} 
\affiliation{Steward Observatory, University of Arizona,
933 N Cherry Ave, Tucson, AZ, 85721-0009, USA}

\author[0000-0001-7592-7714]{Haojing Yan} 
\affiliation{Department of Physics and Astronomy, University of Missouri,
Columbia, MO 65211, USA}

\author{Min S.~Yun}
\affiliation{Department of Astronomy, University of Massachusetts at Amherst, Amherst, MA 01003, USA}

\author[0000-0002-0350-4488]{Adi Zitrin}
\affiliation{Physics Department, Ben-Gurion University of the Negev, P. O. Box 653, Be’er-Sheva, 8410501, Israel}

\begin{abstract}
A Type~Ia supernova (SN) at $z=1.78$ was discovered in James Webb Space Telescope Near Infrared Camera imaging of the galaxy cluster PLCK G165.7+67.0 (G165; $z = 0.35$). The SN is situated 1.5--2\,kpc from the host-galaxy nucleus and appears in three different locations as a result of gravitational lensing by G165. These data can yield a value for Hubble's constant using time delays from this multiply-imaged SN~Ia that we call ``SN H0pe."  Over the cluster, we identified 21 image multiplicities, confirmed five of them using the Near-Infrared Spectrograph, and constructed a new lens model that gives a total mass within 600\,kpc of ($2.6 \pm 0.3) \times 10^{14}$\,\Msol. 
The photometry uncovered a galaxy overdensity coincident with the SN host galaxy.
NIRSpec confirmed six member galaxies, four of which surround the SN host galaxy with relative velocity $\apll$900 km~s$^{-1}$ and projected physical extent $\lesssim$33 kpc. 
This compact galaxy group is dominated by the SN host galaxy, which has a stellar mass of $(5.0 \pm 0.1) \times 10^{11}$\,{\Msol}.
The group members have specific star-formation rates of  2--260\,Gyr$^{-1}$ derived from the H$\alpha$-line fluxes corrected for stellar absorption, dust extinction, and slit losses. Another group centered on a strongly-lensed dusty star forming galaxy is at $z=2.24$. The total (unobscured and obscured) SFR of this second galaxy group is estimated to be ($\apg$100\,\Msol\,yr$^{-1}$), which translates to a supernova rate of $\sim$1 SNe~yr$^{-1}$, suggesting that regular monitoring of this cluster may yield  additional SNe.
  \end{abstract}

\keywords{large-scale structure of universe - gravitational lensing: strong – galaxies: fundamental parameters – galaxies: clusters: general – galaxies: high-redshift}

\section{Introduction} \label{sec:intro}
A lensing cluster can generate giant arcs which are as spectacular as they are important. For certain fortuitous geometrical arrangements of observer, lens, and source, the image of a single source in the background can be observed at multiple locations on the sky (``the image plane'').  Such a set of multiple images of one source (usually a galaxy) constitutes an ``image system." Because the light paths must trace backwards from the observer through all of an image system's observed positions to a single source, the source's redshift together with the positions and orientations of the lensed images constrain the distribution of the lensing matter, both dark and luminous  \citep[][and references therein]{Kneib2011}. When a source object varies with time, its multiple images also vary but with different time delays depending on the light-travel time along each path to the observer.  \citet{Refsdal1964} first proposed that given a multiply-imaged SN, its redshift, and a lens model, the value for the Hubble's constant, $H_0$, can be inferred from the observed time delays between the images (barring extreme microlensing/millilensing effects; \citealt{Goobar2017, Dhawan2020}). Refdal's idea has come to be known as ``time-delay cosmography.''

\subsection{The G165 Cluster Lensing Field}
One galaxy cluster with ample lensing evidence is PLCK G165.7+67.0 (``G165''). G165 ($z=0.35$) first garnered attention by its gravitational amplification of a single background galaxy, now called Arc~1. This galaxy is boosted to an observed sub-mm flux density $S({350\,\micron})>700$\,mJy \citep{Canameras2015,Harrington2016}, making it detectable by the {Planck} and {Herschel Space Observatory} missions \citep{PlanckCollaboration2016,Planck2020}.
Strong lensing by G165 renders Arc~1 into an image system consisting of two images (Arc 1b \& Arc 1c) that merge with the critical curve and a counter-image that we call ``Arc 1a" hereafter).  The star-formation rate (SFR) for Arcs~1b/1c uncorrected for lensing magnification $\mu$ was estimated as  $\mu\rm SFR \sim 12,000$--24,000\,\Msol\,yr$^{-1}$, based on integrating the spectral energy distribution (SED) fit over a wavelength range of 8--1000\,$\mu$m \citep{Harrington2016}. A requisite byproduct of this high SFR is ultraviolet (UV) radiation from massive stars, which was detected in Arc~1b/1c in the observed-frame $g$-band \citep{Frye2019}, redshifted from the rest UV by $z=2.24$ \citep{Harrington2016}. The physical conditions of this dusty star-forming galaxy \citep[DSFG;][]{Casey2014} are poorly constrained owing to the lack of a rest-frame optical spectrum with which to obtain the galaxy classification \citep[e.g.,][]{Mingozzi2023}, estimate the dust extinction, investigate the star-forming properties via the Balmer emission lines \citep[e.g.,][]{Kennicutt1998,Shapley2022}, and measure the gas-phase oxygen abundance  \citep[e.g.,][]{Curti2020,Li2023}.

\begin{figure}[h]
\includegraphics[scale =0.115]{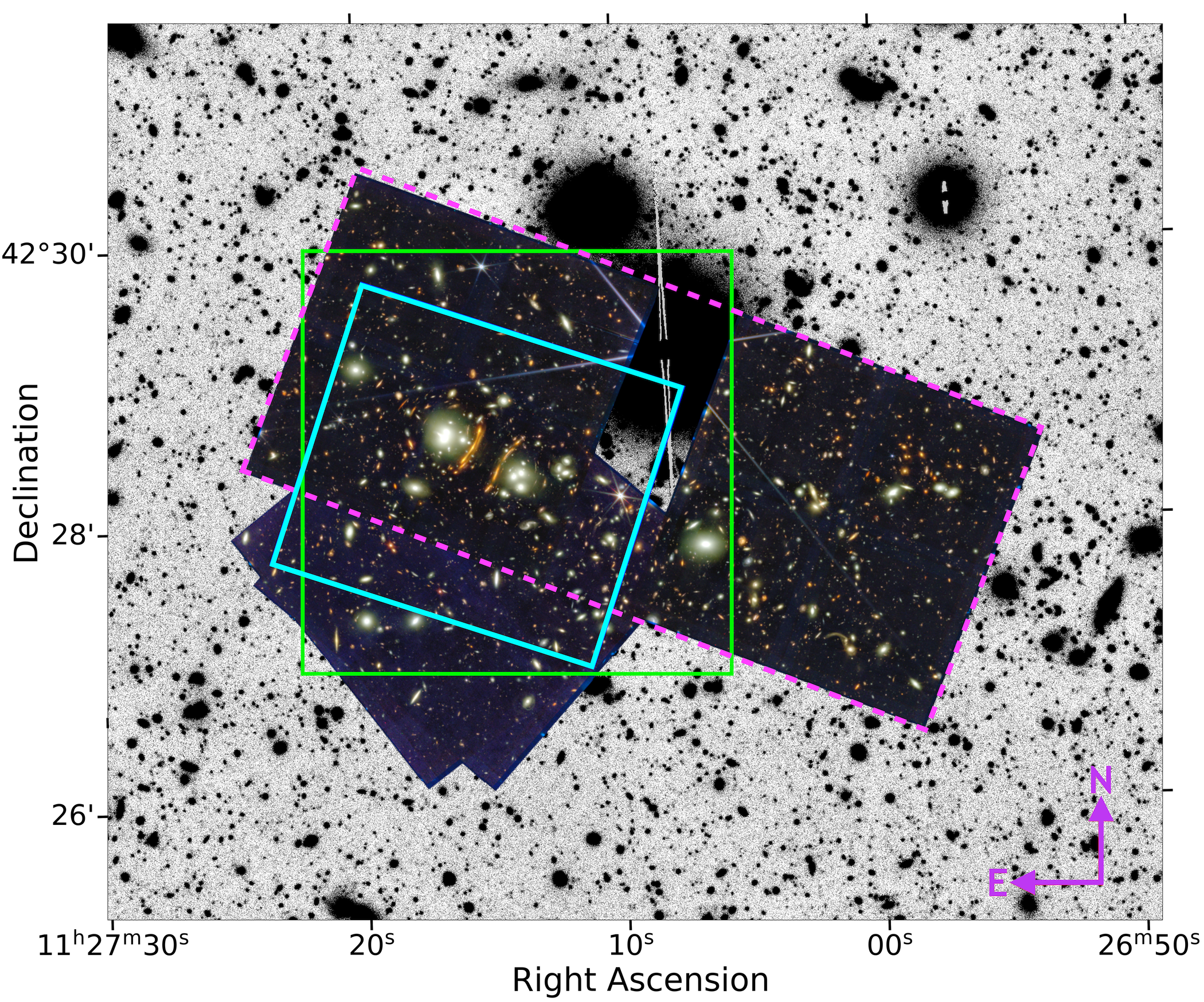}
\caption{JWST/NIRCam coverage of the G165 field.  The background is the $r$-band negative image from CFHT/Megaprime. Superposed color images show the combined NIRCam data. The pink long-dashed rectangle outlines Epoch~1, and Epochs~2 and~3 are squares that mostly overlap each other but have slightly different rotation angles. The blue square outlines the field of view of previous HST WFC3-IR imaging, which usefully covers a portion of the gap between the two NIRCam modules. The green square frames the field of view adopted to construct the lens model.}
\label{fig_foot}
\end{figure}

\citet{Frye2019} used high-resolution HST WFC3-IR imaging to identify \edit1{a total of 11} image  systems in G165.  One of them, called ``Arc~2,'' is also a prominent infrared-bright ($m_K=20.5$ AB) galaxy. Its photometric redshift, estimated from a joint fit of the imaging from seven ground- and space-based facilities \citep{Pascale2022a}, is $\zph=2.30 \pm 0.32$. The similar redshifts of Arc 1 and Arc 2 at $z\approx 2$ taken together with the photometric redshift estimates of several other image families also at $z\approx2$ was somewhat surprising, making it  tempting to suspect that there is a galaxy overdensity at this redshift. However, extant images were of inhomogeneous quality with limiting magnitudes ranging from 23.3--28.9 mag (AB) and point-spread function (PSF) full-width-half-maximum values ranging from 0\farcs13--2\farcs02, complicating  any confirmation of $z\sim2$ sources by their photometric redshifts alone. More recently, \citet{Polletta2023} measured a ground-based spectroscopic redshift for Arc~2 at $\zsp = 1.783 \pm 0.002$. 

In addition to identifying new image systems, the HST WFC3-IR imaging enabled construction of new lens models for G165 \citep[][]{Canameras2018,Frye2019}. These models confirmed that both the northeastern (NE) and southwestern (SW) visible galaxy concentrations of this apparent binary cluster are mass concentrations. The derived total mass was high, ${\sim}2.6\times10^{14}$\,\Msol. Despite using 11 image systems, all models were anchored on the spectroscopic redshift for only one image system, Arc 1, thereby limiting the accuracy of the resulting lens model and its ability to recover the lensed image positions \citep{Johnson2016}. 

\begin{figure*}[tb]
\centering
\includegraphics[scale=0.48]{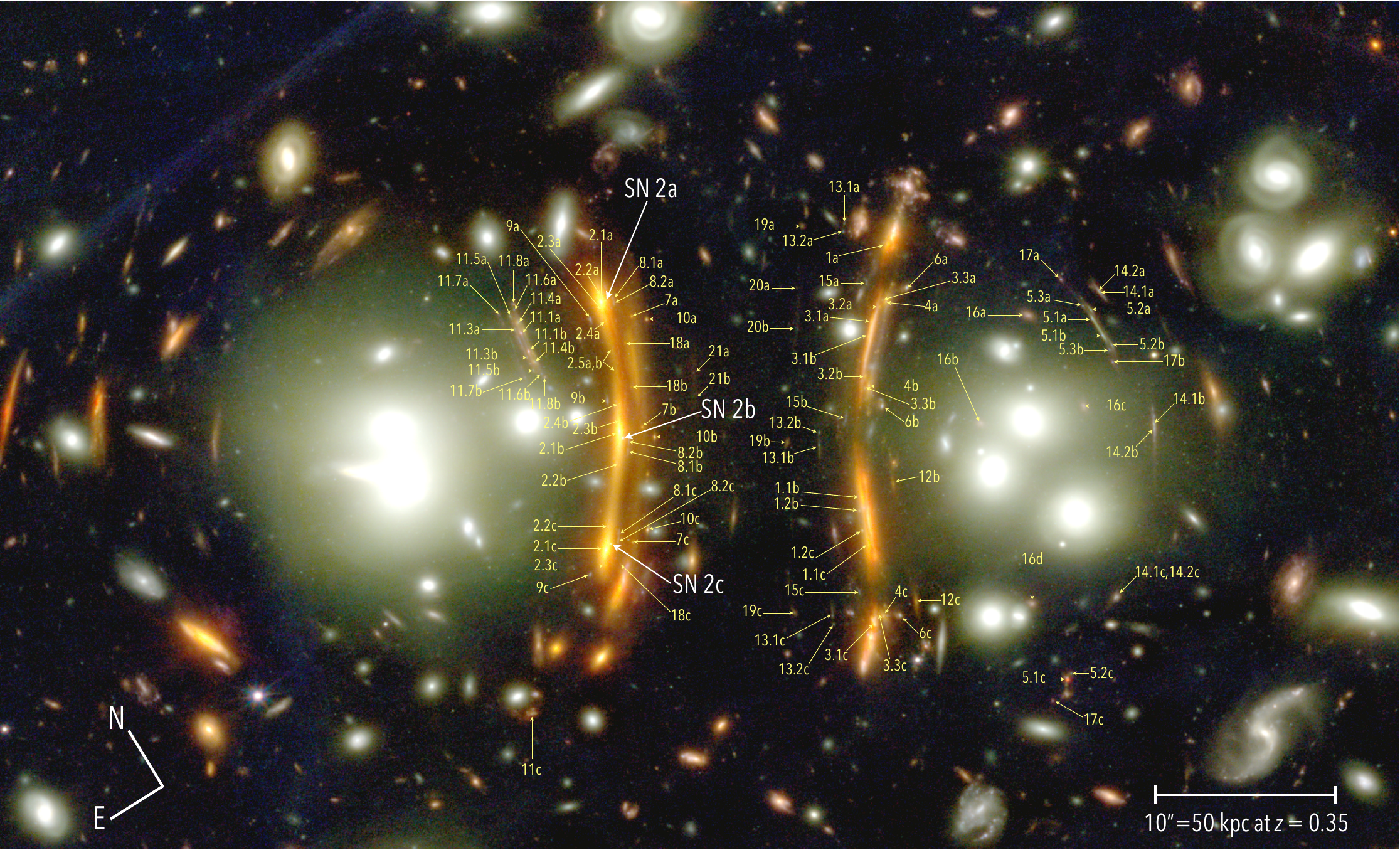}
\caption{ 
NIRCam color composite image of the central region of G165\null. G165 is a double cluster with prominent NE and SW components.
Colors follow the prescription in Trilogy \citep{Coe2012} with red showing F444W and F356W, green showing F277W and F200W, and blue showing F150W and F090W. The 21 image systems used in our lens model are labeled. They include the DSFG as Arcs~1a and 1b/1c. The triply-imaged SN Ia is labeled as ``SN 2a/2b/2c."  The orientation and image scale are provided for reference. All known arc substructures are marked, even those not used in the lens model. 
} 
\label{fig_map}
\end{figure*}

\subsection{Motivation to Search for Multiply-imaged Supernovae}\label{sec:transients}

One of the brightest commonly-occurring transients with a relatively short timescale in a galaxy cluster field is a SN\null. Measuring a difference in the time of peak brightness in the light curve of each image from a multiply-imaged SN enables a robust measurement of the relative time delay between the images.  The time delay(s) plus an accurate lens model and values for the lens and source redshifts give the ratio of the angular diameter distances, which in turn depends directly on $H_0$ \citep{Linder2011, Pierel2021, Treu2022, Suyu2023}.

The first spectroscopically-confirmed multiply-imaged Type Ia supernovae (SNe Ia) were ``iPTF16geu" at $z=0.409$ with an Einstein ring radius $\theta \approx 0\farcs3$ \citep{Goobar2017} and ``SN Zwicky" at $z=0.3554$ with $\theta \approx0\farcs18$  \citep{Goobar2023,Pierel2023}. The time delays for both of these were $\sim$hours with uncertainties of $\sim$1--2 days, and therefore the uncertainties in the distance ratios were large. This somewhat restricted their usefulness as cosmological probes \citep{SainzdeMurieta2023}. Aside from the uncertainties, the SN redshifts $\lesssim$0.4 were in a range where other methods have already constrained $H_0$. The first cluster-lensed SN was ``SN Requiem,"  a photometrically classified SN~Ia  at  $z=1.95$. Its galaxy images are well separated, but the SN was discovered in archival images three years post-event. A long-delay image is expected to appear in 2037 \citep{Rodney2021}, and an $H_0$ measurement from SN Requiem will have to wait until then. The most recent multiply-imaged SN  2022riv\footnote{\url{https://www.wis-tns.org/object/2022riv}} was discovered after all but one image had faded. SN 2022riv was observed with {JWST} (JWST-DD-2767) with the aim of detecting the earlier images. While it was confirmed as Type Ia, no other images were detected, precluding an $H_0$ measurement (P.~Kelly, priv.~comm.\ 2023).

By far the best-studied multiply-imaged SN  is ``SN~Refsdal" at $z = 1.49$.  SN~Refsdal appeared in 2014 in four different locations as a result of strong lensing by the galaxy cluster MACS J1149.5+2223 \citep{Kelly2015}.  A fifth image appeared about a year later and $\sim$8\arcsec\ away. The time delay between the two sets of SN images \citep{Kelly2023b} enabled a measurement of $H_0$ \citep{Kelly2023a} with precision near 7\%. SN~Refsdal is the first case for which $H_0$ has been measured by way of a multiply-imaged SN \citep[][and references therein]{Kelly2023b,Kelly2023a}. SN~Refsdal was a Type II SN\null, so there has yet to be a measurement of $H_0$ with a SN Ia.  While a time-delay measurement is possible with any SN type, precision is expected to be higher for SNe Ia because of their well-understood light curves \citep[e.g.,][]{Pierel2019, Pierel2021}. Additionally, the standard-candle nature of SN Ia can allow a rare direct measurement of the absolute magnification of the  lens, which is not possible with other SN types.


\subsection{Outline of this paper}
\label{sec:outline}
On account of G165's physical properties, prominently-placed infrared galaxies,  and ample strong-lensing evidence, the PEARLS {JWST} GTO program (PID \#1176; PI: R.~Windhorst) obtained NIRCam images of the cluster. This imaging, which is referred to as ``Epoch~1,"
uncovered three point sources which were bright ($m_{\rm F150W}=23.91\pm0.01$ AB) in 2023 but not present in 2016 HST imaging.  This triply-imaged transient  was estimated with $>$90\% probability to be a Type~Ia SN based on the three light-curve points obtained in Epoch~1 \citep{Frye2023an}.  Follow-up with a JWST disruptive Director's Discretionary Time (DDT) program (PID: 4446, PI: B.~Frye) provided two additional imaging epochs yielding a total of 9 samplings (one from each of the three images in each observing epoch) of the SN light curve. The cadence bracketed the second SN peak at $\lambda > 1.8\,\mu$m characteristic of SNe Ia. NIRSpec spectroscopy was also obtained for all three SN images during Epoch~2, when they were relatively bright. We refer to this transient as ``SN H0pe" for its potential to measure the time delays between the images and from those to measure $H_0$. 

This paper presents an overview of the initial science results from the combined JWST PEARLS and DDT imaging and from spectroscopic observations in the G165 cluster field, including the discovery and early analysis of SN H0pe. 
This study is the first in a series of papers whose objective is to investigate SN H0pe, the cluster, and the lensed sources. 
This paper is organized as follows. \S\ref{sec:obs} introduces the JWST and ancillary data sets. \S\ref{sec:phot} describes the NIRCam photometry, the estimates of photometric redshifts, and the photometric discovery of SN H0pe. The construction of the lens model appears in \S\ref{sec:LTM}. The NIRSpec spectroscopic analysis follows in \S\ref{sec:analysis} with a focus on the Arc~1 and Arc~2 $z\approx 2$ galaxy groups.  The physical properties of G165 and of the high-redshift galaxy groups are investigated in \S\ref{sec:disc}, and \S\ref{sec:finis} summarizes the results. This paper uses the AB magnitude system throughout, and redshift distances are based on a  flat $\Lambda$CDM cosmology with $H_0=67$~km~s$^{-1}$\,Mpc$^{-1}$,  $\Omega_{m,0}=0.32$, and $\Omega_{\Lambda,0}=0.68$ \citep{PlanckCollaboration2018}.

\medskip

\section{Observations and Reductions} \label{sec:obs}
\subsection{JWST/NIRCam} \label{sec:NIRCam} 
The Epoch 1 NIRCam observations were obtained as part of the PEARLS {\it JWST} GTO program.
The observing date was selected to minimize stray light expected from a nearby bright star. Exposures were taken in four filters in the short wavelength (SW) channel and four in the long wavelength (LW) channel as shown in Table~\ref{tab_exposure}. Both NIRCam modules collected data. Epochs 2 and 3 of NIRCam imaging were acquired as part of the JWST disruptive DDT program (PID 4446, PI: Frye) to follow the supernova's light curve in each of its three images. In this follow-up program (also summarized in Table~\ref{tab_exposure}), exposures were taken in six filters using only Module~B of NIRCam.
The NIRCam observations covered the central region of the cluster including both the NE and SW cluster components, the three images of the SN, the DSFG, all of the image systems, and other prominent giant arcs. Figure~\ref{fig_foot} depicts the field coverage overlaid on an extant $r$-band image using the Canada France Hawaii Telescope (CFHT) Megaprime imager.

The NIRCam images were reduced by our team as described by \citet{Windhorst2023}.  Briefly, the data were retrieved from the {Mikulski} Archive for Space Telescopes (MAST), and the latest photometric calibration files were used (pmap\_1100). All images were reduced using version 1.11.2 of the STScI JWST Pipeline \citep{pipeline} with an additional correction for $1/f$ noise by applying the prescription of C.\ Willott.\footnote{\url{https://github.com/chriswillott/jwst.git}}  The ProFound code \citep{Robotham2018} was run, which makes a second round of corrections of other image artifacts in the relevant rows and columns. This step additionally flattens the background and corrects for detector-level offsets, ``wisps," and ``snowballs" \citep{Robotham2017,Robotham2018}. Since the \citet{Windhorst2023} publication, improvements in the data reduction techniques have been made by \citet{Robotham2023} regarding the removal of image wisps by using the wisp-free LW images as priors to identify the outer contours of the real detected objects. Those real objects were subsequently removed from the SW images to get a pure wisp image which was then fully subtracted. This process yields an image noise in the final mosaic that is almost the same in the wisp-removed area as in the surrounding wisp-free areas. 

After each frame was calibrated, the frames were aligned onto a common astrometric reference frame and drizzled into mosaics with pixel scale 20 milli-arcseconds (mas). The process was similar to that first described by \cite{Koekemoer2011} but updated to use the JWST pipeline.\footnote{\url{https://github.com/spacetelescope/jwst}}  
Mosaics were produced for each filter in each separate epoch. For the six filters in common, all epochs were also combined into a grand mosaic for each filter. All mosaics were aligned onto the same pixel grid based on deep, ground-based CFHT/Megaprime images with good seeing on 2014 May 29 (PI: Nesvadba).  The image mosaic was aligned directly onto Gaia DR3 \citep{GAIA2016, GAIA_DR3} by M.~Nonino (priv.~comm.~2023). The NIRCam data were aligned onto this grid with residual RMS below 2--3\,mas and no significant large-scale distortions.
Figure~\ref{fig_map} shows the central region of G165 in the main NIRCam mosaic. 

\begin{deluxetable}{cccc}
\tabletypesize{\footnotesize}
\tablecaption{JWST Epochs and NIRCam Exposure Times}
\label{tab_exposure}
\tablecolumns{4}
\tablehead{
\colhead{\bf Filter} &   \colhead{{\bf Epoch 1}} & \colhead{{\bf Epoch 2}} & \colhead{{\bf Epoch 3}} \\[-2ex]
& \colhead{Mar 30}& \colhead{Apr 22}& \colhead{May 09}
}
\startdata
F090W & 2491 & 1246 & 1417 \\
F115W & 2491 & \no & \no \\
F150W & 1890 & 859 & 1246 \\
F200W & 2104 & 1761 & 1761 \\
F277W & 2104 &1761 & 1761\\
F356W & 1890 & 859 &1246\\
F410M & 2491 & \no & \no \\
F444W & 2491 &1246 &1417 \\
\enddata
\tablecomments{Dates of each epoch are in 2023, and exposure times for each filter are in seconds. The NIRSpec observations (Section~\ref{sec:obsNIRSpec}) were obtained during the Epoch~2 visit.}
\end{deluxetable}

\subsection{JWST/NIRSpec} \label{sec:obsNIRSpec} 
NIRSpec medium-resolution Micro-Shutter Array (MSA) spectroscopy of the G165 field was obtained on 2023 Apr 22 as part of the {JWST} DDT program (PID 4446, PI: Frye). The MSA mask was populated with the positions of the three SN appearances (SN 2a, 2b, and 2c), two of the three images of the SN host galaxy (Arc 2a and 2c), and counterimages of three other image systems (Arcs 5a, 8c, and 9c). The remainder of the mask was filled with other lensed sources which summed to a total of 42 lensed targets. The observations used the grating/filter combinations G140M/F100LP to cover spectral range 0.97--1.84~$\mu$m (rest-frame 0.35--0.66~$\mu$m at $z=1.8$) and G235M/F170LP to cover 1.66--3.17~$\mu$m (rest-frame 0.57--1.1~$\mu$m at $z=1.8$), both with spectral resolution $R\approx1000$. We also acquired a PRISM/CLEAR spectrum covering  0.7--5.3~$\mu$m (rest-frame 0.25--1.9~$\mu$m) with $R \approx 20$--300 (50--14~\AA). All of the seven supplied guide stars were acquired, resulting in especially tight pointing residuals of 1--7\,mas and successful pointing even for targets near the edges of an MSA array. The science exposure times were 4420~s, 6696~s, and 919~s for G140M/F100LP, G235M/F170LP, and the PRISM/CLEAR observations, respectively.  A 3-point nod pattern was selected for each observation, and each MSA slit consisted of 3 microshutters giving slit height 1\farcs52. MSA slits are 0\farcs20 wide in the dispersion direction, and the long dimension was oriented at position angle 276\arcdeg.

The Stage 1 calibrated data were retrieved from MAST and reduced using the JWST NIRSpec pipeline, version 1.11.3.\footnote{\url{https://jwst-pipeline.readthedocs.io/en/latest/index.html}} Stage 2 and 3 reduction used the JWST pipeline with reference files ``jwst\_1100.pmap" for all levels of the data reduction with an exception regarding the background subtraction for extended sources, as described below. Saturated pixels and other image artifacts were flagged in the 2D spectra.  The NIRSpec IRS2 detector readout mode was used, which largely reduced the $1/f$ noise.
The 2D spectra were wavelength- and flux-calibrated based on the calibration reference data system (CRDS) context.  Finally, individual calibrated 2D spectra exposures were coadded, and one-dimensional (1D) spectra plus uncertainties were optimally extracted \citep{Horne1986}.

The pipeline background subtraction performed well for single point sources and single small sources which were fully covered by the aperture. This is because the dithered exposures provided a good ``best-fit" background consisting of the intracluster light and/or other underlying extended sources and/or detector \edit1{offsets}. Hence, the resulting NIRSpec flux from the pipeline directly gave the flux for the point/small source. 

However, the observations did not include exposures of a separate background field. This made it more of a challenge to estimate the background for sources extending across multiple microshutters. One example is the SN host galaxy Arc 2 and the SN, for which all three microshutters are occupied by sources.  For this case, the background template formed in the NIRSpec pipeline comes from the flux through the source shutter in the dithered exposure. This ``image from image" background might include some flux from the galaxy, leading to an oversubtraction of the background. A complementary problem is the case for which neighboring microshutters are occupied by different sources. In the ``MOS Optimal Spectral Extraction” tool from STScI (based on the method from \citealt{Horne1986}) a source kernel and a polynomial background template are fit at the same time for one source within an MSA slit, based on a spatial window in the 2D spectrum chosen manually by the user, but the software does not support multiple source extraction.

To alleviate some of these issues, a custom-built code was developed to perform the background subtraction.
The code is different from the pipeline in that it builds a more locally-derived background template. For each pixel, we evaluated the minimum flux of the set of five dithered pixels. Then for each pixel within each spatial column {$i$}, the best value for the background was computed by the median value of this minimum flux within a running boxcar 10 spatial columns wide and centered on column {$i$}. We found 10 columns to be a good compromise between a smaller median filter starting to encroach on the size of a typical cosmic ray mask and a larger median filter smoothing out the background features in this wavelength-dependent operation. To cope with image crowding, the code has a multiple source extraction mode that fits multiple source kernels simultaneously for each object along the MSA slit.

Operationally, we ran NIRSpec Stage 2 with the background subtraction task turned off and then applied the custom-built code.  The detailed content of this code and its implementation for this data set appear elsewhere \citep{Chen2023H0pe}.  

\begin{figure}[h]
\centering\includegraphics[scale=0.75]{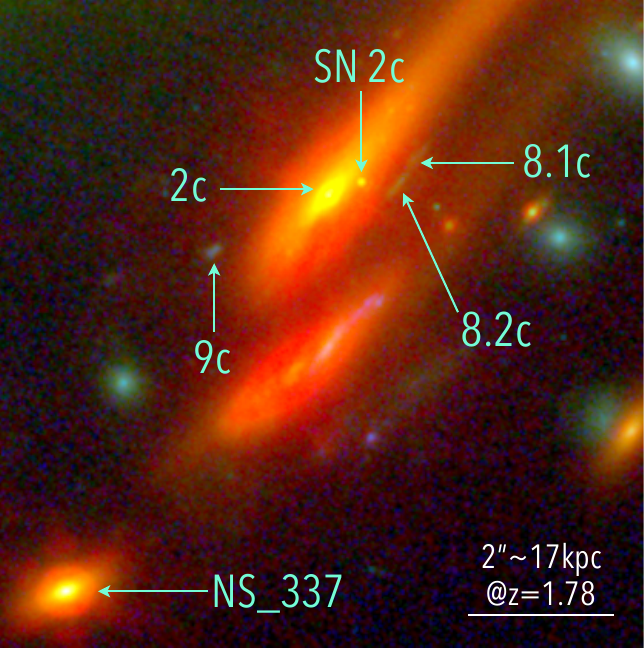} 
\caption{NIRCam color image centered on one of the SN host images 2c (Arc 2c), along with three other galaxies spectroscopically-confirmed to be at $z = 1.78$ in this study. Our lens model predicts for Arcs 2, 8, 9, and NS\_337 to be situated within 33 physical kpc in the source plane. The NIRSpec spectra of these arcs appear in Figure~\ref{fig_specstack}. Colors follow the prescription in Trilogy \citep{Coe2012} with red showing F444W and F356W, green showing F277W and F200W, and blue showing F150W and F090W. 
}
\label{fig_specstackimage}
\end{figure}

\begin{figure*}[tb]
\centering\includegraphics[scale=0.90]{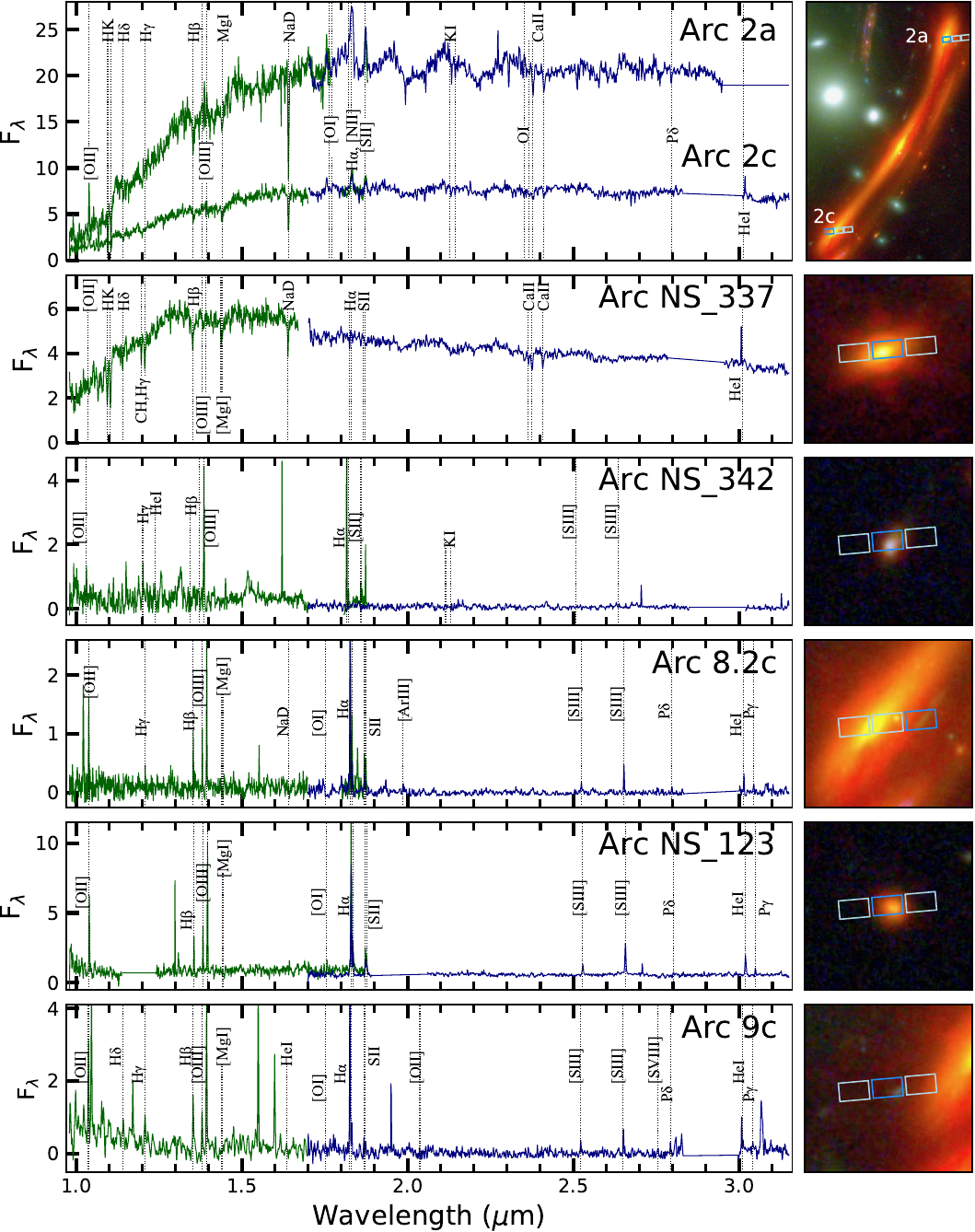}
\caption{NIRSpec spectra of lensed sources at $z\approx1.78$. Wavelengths are in the observed frame, and the ordinate shows $F_{\lambda}$ in units of 
$10^{-19}$\,erg\,s$^{-1}$\,cm$^{-2}$\,\AA$^{-1}$. The G140M spectrum is plotted in green and the G235M spectrum in blue.  Detected lines are marked. The images to the right of each panel show the respective source with MSA slit positions overlaid and are oriented north up, east left. The color rendering is the same as Figure~\ref{fig_specstackimage}. 
The microshutter depicted in blue is the one from which the spectrum was extracted. The spectra are presented in order of  star formation activity with more quiescent sources with weaker H$\alpha$ emission lines and stronger 4000\,\AA\ and Balmer breaks at the top  to emission-line sources with multiple nebular emission lines  at the bottom. These six sources uncover a diverse set of galaxy properties all contained in this single high-redshift galaxy overdensity. The spectrum for Arc~2a appears brighter and redder than the one for Arc~2c owing to the slit being better centered on the source. 
}
\label{fig_specstack}
\end{figure*}

In all, the NIRSpec spectroscopy produced a total of 47 1D spectra.
We measured the redshifts from emission- and absorption-line features of each source, as available. The line centers were determined by fitting Gaussians to each spectroscopic line feature using {\sc specutils} \citep{specutils}. Of 47 spectra, 30 produced secure redshifts, which are listed in Table~\ref{tab_NIRSpec}. A redshift is considered secure if it has a high-significance detection of two or more spectral features and ${>}2\sigma$ level in the continuum. Of these, the highest-redshift galaxy is NS\_274, a relatively rare example of a quiescent galaxy for which we measure $\zsp = 4.1076 \pm 0.0023$. The highest redshift multiply imaged galaxy is Arc 5.1a with a redshift measured from Balmer lines from H$\alpha$ through H$\epsilon$ detected in emission of $\zsp=3.9530 \pm 0.0004$. The spectroscopic analysis of the NIRSpec spectra of galaxy images at $z=1.78$ and $z=2.24$ appear in \S\ref{sec:analysis}. (The prefix ``NS" stands for NIRSpec, and it precedes all of the NIRSpec-confirmed galaxy images in this study in Table~\ref{tab_NIRSpec}.)

The three SN~H0pe spectra are of high quality. Most prominent is the requisite detection of the \ion{Si}{2}\,$\lambda$6355 absorption feature blueshifted to $\sim$6150\,\AA, closely followed by the detection of the [\ion{Ca}{2}]\,$\lambda\lambda$8498,8542,8662 IR triplet ``CaT," amongst other spectroscopic features. The spectrum of the SN, the SN classification as Type Ia, and the measurement of the spectroscopic time delay will appear in a different paper \citep{Chen2023H0pe}. The spectra of the SN host galaxy Arcs~2a and 2b had pre-existing redshifts, both based on the joint detection of [\ion{O}{2}]$\,\lambda$3727 and the 4000\,\AA\ and Balmer breaks \citep{Polletta2023}. The new NIRSpec spectra give the first redshift for Arc~2c, whose value matches that of Arc~2a (Table~\ref{tab_NIRSpec}).  Nearly 20 spectroscopic features are detected as well as the 4000\,\AA\ and Balmer breaks. Somewhat remarkably, six different lensed sources have the same redshift as the SN\null. Their images are shown in Figure~\ref{fig_specstackimage}, and their spectra are presented in Figure~\ref{fig_specstack}. 
This redshift is interesting because it also coincides with the strongest peak in the photometric redshift distribution after the cluster redshift, as described in \S\ref{sec:phot_zph}.

\begin{figure*}[tb]
\centering\includegraphics[scale=0.90]{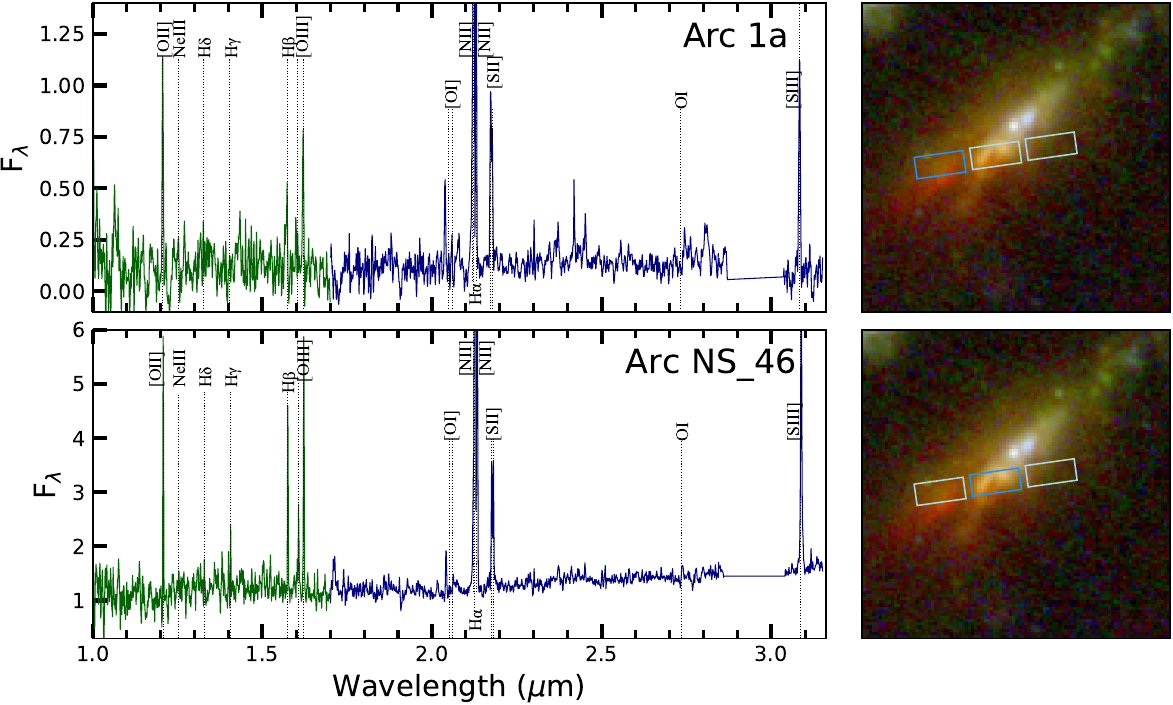}
\caption{NIRSpec spectra of Arc~1a (=Arc~NS\_969, top) and a nearby object (Arc~NS\_46, bottom). Wavelengths are in the observed frame, and the spectra are in $F_{\lambda}$ in units of 
$10^{-19}$\,erg\,s$^{-1}$\,cm$^{-2}$\,\AA$^{-1}$. 
The G140M spectrum is plotted in green and the G235M spectrum in blue. Detected lines are marked. The images to the right of each panel show the respective source with MSA slit positions overlaid, and are oriented north up, east left. 
The color rendering is the same as Figure~\ref{fig_specstackimage}. The objects were observed in the same triplet of MSA slits with Arc~1a in the left segment and Arc~NS\_46 in the middle segment, neither perfectly centered in the respective segments.  The blue outline in each image shows the microshutter from which the spectrum to its left was extracted.
}
\label{fig_specDSFG}
\end{figure*}

Arc 1b/1c had a previously measured  $\zsp = 2.2357 \pm 0.0002$ \citep{Harrington2016}. The NIRSpec spectrum (NS\_969) gives the first redshift of its counterimage, Arc~1a, and the redshifts agree. A second spectrum (NS\_46) in an adjacent MSA microshutter (0\farcs46 northwest) is redshifted by 
$\sim$1400\,km\,s$^{-1}$ relative to Arc~1a. Both spectra, shown in Figure~\ref{fig_specDSFG}, have strong nebular emission lines and starburst attributes. Section~\ref{sec:anal_Arc1} describes their nebular and stellar properties.  

In some cases, most notably for Arcs 2a, 2c, and Arc~NS\_337, aperture losses are expected because the MSA slit coverage is smaller than the source size. To account for this shortfall, for each filter bandpass, synthetic photometry was computed based on the NIRSpec PRISM spectra, which provided continuous coverage over the wavelength range of all eight NIRCam bands.  We then compared our results to the NIRCam photometry integrated over the entire source. We refer to \S\ref{sec:analysis} for details.

\begin{deluxetable*}{cccCC}
\tabletypesize{\footnotesize}
\tablecaption{NIRSpec Spectra}
\tablecolumns{5}
\tablehead{
 \colhead{{\bf ID}} & \colhead{{\bf R.A.}} &  \colhead{{\bf Decl.}} & \colhead{\textbf{$m_{\rm F200W,obs}$}} & \colhead{\zsp} 
}
\startdata
NS\_2 (SNa) & 11:27:15.31 & +42:28:41.02 & \nodata$^a$ & \nodata$^b$ \\
NS\_3 (SNb) & 11:27:15.60 & +42:28:33.73 & \nodata$^a$ & \nodata$^b$ \\
NS\_4 (SNc) & 11:27:15.94 & +42:28:28.90 & \nodata$^a$ & \nodata$^b$ \\
NS\_6 (Arc 2c) & 11:27:15.98 & +42:28:28.72 & 20.26 & 1.7834\pm 0.0005\rlap{\ensuremath{^c}} \\
NS\_7 (Arc 2a) & 11:27:15.34 & +42:28:41.05 & 20.30 & 1.7833\pm 0.0010\rlap{\ensuremath{^c}} \\
NS\_19 (Arc 5a)& 11:27:13.20 & +42:28:25.73 & 25.36 & 3.9530\pm 0.0004 \\
NS\_26 (Arc 8.2c)& 11:27:15.89 & +42:28:28.90 & 26.13 &1.7839\pm 0.0002\\
NS\_29 (Arc 9c)& 11:27:16.11 & +42:28:27.99  & 26.10 &1.7816\pm 0.0002 \\
NS\_46 & 11:27:13.87 & +42:28:35.67 & 24.04 & 2.2401\pm0.0002  \\
NS\_104 & 11:27:00.84 & +42:27:03.65 & 24.83 & 3.1109\pm0.0004 \\
NS\_112 & 11:27:02.46 & +42:27:10.22 & 21.59 & 0.6227\pm 0.0003\\
NS\_123 & 11:27:04.64 & +42:27:15.24 & 23.94 &1.7874\pm0.0003 \\
NS\_143 & 11:27:07.64 & +42:27:26.22 & 23.28 & 1.6322\pm 0.0002 \\
NS\_171 & 11:27:05.82 & +42:27:37.10 & 23.33 & 1.1787\pm 0.0002 \\
NS\_274 & 11:27:11.67 & +42:28:10.70 & 23.44 & 4.1076\pm0.0023 \\
NS\_285 & 11:27:06.82 & +42:28:12.88  & 23.81 & 0.4466\pm 0.0001\\
NS\_337 & 11:27:16.28 & +42:28:23.59 & 21.77 & 1.7810\pm0.0009\\
NS\_342 & 11:27:19.46 & +42:28:24.79 & 24.06 & 1.7664\pm0.0009 \\
NS\_376 & 11:27:16.91 & +42:28:30.62 & 22.69 & 0.6840\pm 0.0001\\
NS\_407 & 11:27:06.62 & +42:28:37.80 & 22.49 & 1.8553\pm0.0009\\
NS\_411 & 11:27:05.17 & +42:28:37.75 & 24.28 & 1.2456\pm 0.0007 \\
NS\_477 & 11:27:04.23 & +42:28:56.36 &  23.00 &1.8524\pm0.0009 \\
NS\_481 & 11:27:13.66 & +42:28:57.12 & 24.35 & 1.3236\pm 0.0002 \\
NS\_505 & 11:27:18.95 & +42:29:06.64 & 24.40 & 1.2888\pm 0.0002 \\
NS\_511 & 11:27:18.69 & +42:29:08.42 & 24.13 & 3.3255\pm 0.0006 \\
NS\_548 & 11:27:13.21 & +42:29:30.75 & 23.14 & 0.8161\pm 0.0003 \\
NS\_610 & 11:27:00.97 & +42:26:48.25 & 22.77 & 0.6628\pm 0.0004 \\
NS\_969 (Arc 1a)& 11:27:13.92 & +42:28:35.43 & 24.90 & 2.2355\pm 0.0003 \\
NS\_1115 & 11:27:03.09 & +42:29:07.03 & 23.61 & 2.0585\pm 0.0007 \\
NS\_1477 & 11:27:16.65 & +42:27:23.57 & 21.04 & 0.7203\pm 0.0003 \\
\enddata
\tablecomments{NS numbers refer to the MSA slit identifications assigned when the observations were designed. Positions are object positions as measured on NIRCam images.}
\tablenotetext{a}{The photometry is presented by \citet{Pierel2023H0pe}}
\tablenotetext{b}{This spectroscopic redshift is presented by \citet{Chen2023H0pe}.}
\tablenotetext{c}{\citet{Polletta2023} used LBT/LUCI spectra covering observed wavelengths 950--1370\,nm to measure $z=1.782$ for Arc~2a and $z= 1.783\pm0.002$ for the average of Arcs 2a and 2b, fully consistent with the NIRSpec redshifts.} \label{tab_NIRSpec}
\end{deluxetable*}

\begin{figure}[h]
\centering\includegraphics[scale =0.75]{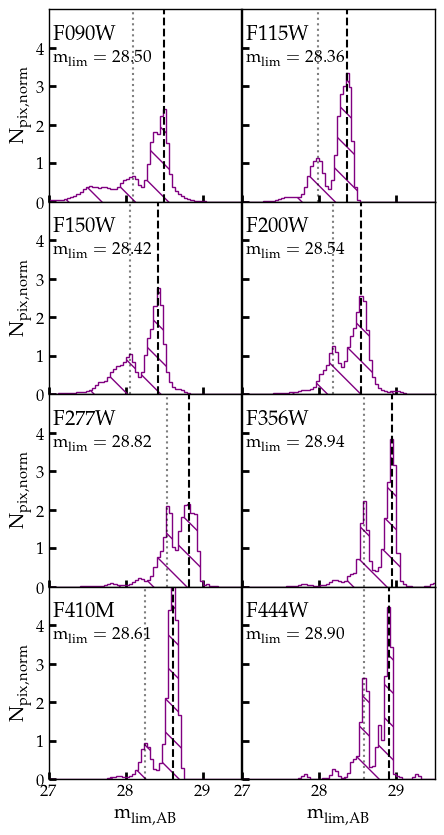}
\caption{
Limiting magnitudes ($5\sigma$) within a 0\farcs2 aperture based on the corrected RMS maps  (\S\ref{sec:phot_NIRCam}).  The dashed vertical lines mark the mode of each distribution, and values are given in each panel's label. The dotted line marks the shallower peak, corresponding to the non-overlapping regions of Epochs~2 and~3. The deeper peak corresponds to the non-overlapping regions of the Epoch~1 images, however the dithering causes the edges of the Epoch~1 images to contribute to the lower peak due to the lower exposure time in those regions. The tail of the deeper peak towards still fainter magnitudes results from the combination of the three epochs of imaging. It is absent in the F115W and F410M filters, which were acquired only in Epoch~1.}
\label{fig_maglimit}
\end{figure}

\subsection{Ancillary Imaging}
 \label{sec:Ancillary}
Other imaging exists to complement the NIRCam imaging. LBT Large Binocular Camera (LBC) $gi$-band imaging was acquired on 2018 January 20 (2018A; PI: Frye). Images reach 3$\sigma$ limiting magnitudes of 25.42 and 24.67 for $g$ and $i$, respectively \citep{Pascale2022a}. In the near-infrared, HST WFC3-IR exposures were taken on 2016 May 30. Images at F110W and F160W reach 3$\sigma$ limiting magnitudes of 28.94 and 27.97 AB mag, respectively (Cy23, GO-14223, PI: Frye).  Details of the observations, reduction, and analysis were given by \citet{Frye2019}.  

At longer wavelengths, LBT LUCI-ARGOS imaging was obtained in $K$ (2016B; PI: Frye).  LUCI+ARGOS corrects the atmosphere for ground-layer distortions via multiple artificial stars that are projected by laser beams mounted on each of the two 8.4\,m apertures \citep{Rabien2019}. This imaging achieved a mean point-source FWHM of 0\farcs29 but a limiting magnitude of 24.07 \citep{Frye2019}. And finally, Spitzer Space Telescope ({Spitzer}) Infrared Array Camera (IRAC) 3.6 and 4.5\,$\mu$m images were acquired in 2017 (Cy 13, PID 13024, PI: Yan).  The {HST}, LBT LUCI+ARGOS $K$-band observations, and the {Spitzer} observations provide baseline imaging prior to the SN event but otherwise are superseded by the deeper and higher-resolution NIRCam images. 

\subsection{Ancillary Spectroscopy}
\label{sec:anc_spec}
The redshifts of image systems and of cluster members are most relevant to this study. This is because they provide inputs to the lens model, whereas singly-imaged or foreground galaxies do not. The NIRSpec observations (\S\ref{sec:obsNIRSpec}) contributed the  redshifts of image systems. (The Arc~2 redshift was already known from LBT/LUCI observations \citealt{Polletta2023}.) 
Redshifts of cluster members come from a MMT/Binospec spectroscopy program and from the literature and are described below.

MMT/Binospec long-slit spectroscopy was obtained in DDT on 2023 Apr 18 (PI: Willner), near in time to the Epoch~2 {JWST} observations of SN~H0pe.  The objective was to confirm the redshift of the brightest cluster galaxy (BCG), which resides in the NE cluster component. The slit position angle 118\fdg5 was selected to intercept the BCG  and another cluster member in  the southwest component that lacked a spectroscopic redshift. The data were acquired with the G270 grating to provide wavelength coverage from 3800 to 9200\,\AA. 
The total science exposure time was 1800~s. The observing conditions were good and reasonably stable, with atmospheric seeing ranging from 1\farcs17--1\farcs25. 

The data were reduced using the observatory pipeline, which performed the usual calibrations (bias correction, flat-fielding, wavelength calibration and relative flux correction), as well as the coaddition of the three separate 600\,s exposures and extraction of the 1D spectra.\footnote{https://bitbucket.org/chil\_sai/binospec} Although redshift-fitting software was available, we opted to measure the spectroscopic redshifts using our own software that contains a library of spectroscopic features and a reference sky spectrum extracted from the pre-sky-subtracted data. Table~\ref{tab_other_spectra} gives positions and redshifts for galaxies intercepted by the long slit, including the BCG at $z=0.3368$, consistent with the previous value from the Sloan Digital Sky Survey DR13 archives.\footnote{Funding for the Sloan Digital Sky Survey IV has been provided by the Alfred P.\ Sloan Foundation, the U.S. Department of Energy Office of Science, and the Participating Institutions. SDSS acknowledges support and resources from the Center for High-Performance Computing at the University of Utah. The SDSS-IV web site is \url{www.sdss4.org}.}
The other five redshifts are new, but only one galaxy is a cluster member. Thus this MMT program contributed the redshift of one new cluster member.

Spectroscopic redshifts were also drawn from the literature. 
\citet[][and references therein]{Pascale2022a} provided 273 redshifts, all obtained prior to the {JWST} observations of SN~H0pe. 
Of those, 34 are within $\pm$4000\,km\,s$^{-1}$ of the cluster mean redshift  $z=0.348$ and within a projected radius of 1~Mpc of the cluster center. We consider these 34 objects to be confirmed cluster members.

In all, the ancillary spectroscopy contributed one cluster member from MMT/Binospec that is new to this study and 34 redshifts of cluster members from the literature. The total sums up to 35 redshifts of cluster members. These galaxies were supplemented by the photometrically-selected counterparts to make a main catalog of cluster members that is described in \S\ref{sec:LTM}.

\section{Imaging Results} \label{sec:phot}

\subsection{NIRCam Photometry} \label{sec:phot_NIRCam}

The extraction of multi-band photometry broadly followed the approaches of \cite{Merlin2022} and \cite{Paris2023}, which balance the need to make faint image detections but also to limit the introduction of spurious sources.  
Initial source detection was performed using SExtractor \citep{Bertin1996} in a two-step HOT+COLD process similar to \citet{Galametz2013}. The JWST NIRCam F200W image, which corresponds to the diffraction limit of the telescope, was assigned as the reference image. For severely blended objects, a separate catalog using F090W for detection was introduced owing to the sharper PSF. 

The object fluxes and uncertainties were measured in each filter using {\sc aphot} \citep{Merlin2019} by assigning Kron-like elliptical apertures, isophotal apertures, and circular apertures of diameter 0\farcs3. To compute realistic photometric uncertainties, 5000 point sources from \texttt{WebbPSF} \citep{Perrin2015} of known fluxes were injected into blank regions of the images, and fluxes and uncertainties were estimated using 0\farcs1 apertures using {\sc aphot} with the RMS maps associated with the images. The RMS maps were then rescaled such that the RMS of the measured flux distribution was consistent with the values expected from {\sc aphot}\null. Because the mosaic was drizzled from 3 separate pointings, this analysis was performed separately on each overlapping and non-overlapping region between the 3 epochs of data.

The final photometry is PSF-matched using PSF models generated from \texttt{WebbPSF}, where all filters are degraded to the PSF of the F444W image using \texttt{pypher}. The F444W image was chosen because it has the largest PSF\null. The PSF models provided satisfactory convolution kernels and ameliorate the trend for the $\rm F200W-F444W$ colors to be bluer than their true values for multi-exposure image mosaics simulated for the CEERS project \citep{Bagley2022}.  \citet{Pascale2022a} gave details of the PSF-matching implementation.

The histograms of the limiting magnitudes  are shown for each filter in Figure~\ref{fig_maglimit}. 
The double-peaked distributions demonstrate the multi-epoch experimental setup, with the PEARLS (Epoch 1) being deeper, and so showing a peak at fainter limiting magnitudes. The tail of the Epoch 1 peak towards fainter magnitude emerges from the combination of all three epochs and is not present in the F115W and F410M filters, which have coverage only in Epoch~1.
The data are shallowest in F115W ($m_{\rm lim} = 28.34$~AB) and are deeper in all of the LW filters relative to the SW filters.

\begin{deluxetable}{ccccc}
\tabletypesize{\footnotesize}
\tablecaption{MMT/Binospec Spectroscopy}
\tablecolumns{5}
\tablehead{
 \colhead{{\bf ID}} & \colhead{{\bf R.A.}} & \colhead{{\bf Decl.}} & \colhead{\textbf{$m_{\rm obs}$$^a$}} & \colhead{\zsp}  
}
\startdata
BCG  & 11:27:16.70 & +42:28:38.75 & 17.69 & 0.3368 \\ 
s101 & 11:27:14.05 & +42:28:21.12 & 18.81 & 0.3427 \\ 
s102 & 11:27:08.04 & +42:27:44.95 & 21.52 & 0.2757 \\ 
s103 & 11:27:20.51 & +42:29:02.01 & 22.14 & 0.4135 \\ 
s104 & 11:27:35.83 & +42:30:33.25 & \nodata${^b}$ & 0.2300 \\ 
s105 & 11:27:32.70 & +42:30:14.07 & \nodata${^b}$ & 0.3509 \\ 
\enddata
\tablecomments{The object identifications (IDs) are arbitrary and used only for convenience in this paper. Positions are object positions as measured on NIRCam images.}
\tablenotetext{a}{The AB magnitude is measured in the F090W filter that most overlaps with the MMT/Binospec spectroscopic coverage.}
\tablenotetext{b}{Source is outside the field of view of the JWST PEARLS program.}
\label{tab_other_spectra}
\end{deluxetable}

\subsection{Photometric Redshift Estimates} \label{sec:phot_zph}

\begin{figure}[h]
\centering\includegraphics[scale=0.45]{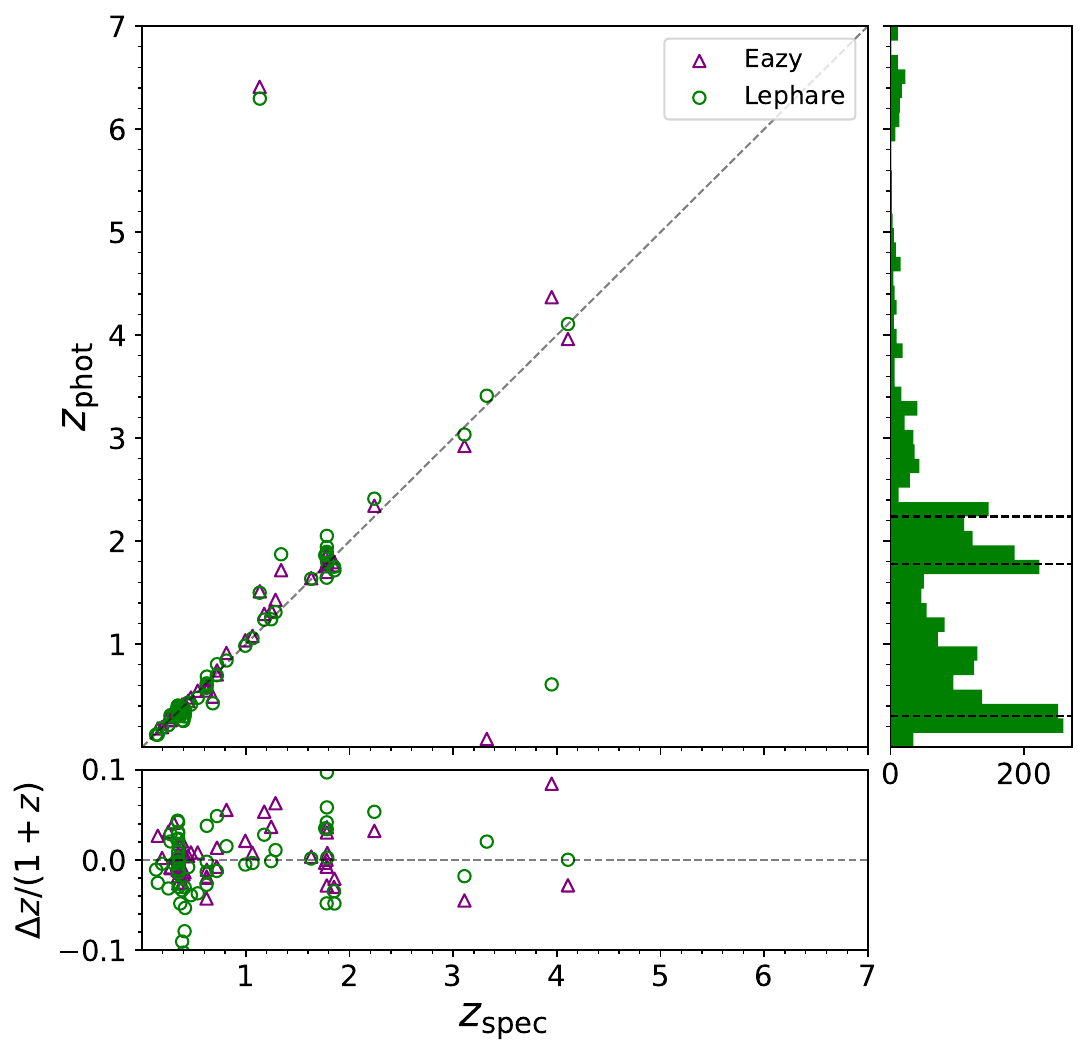}
\caption{Photometric vs.~spectroscopic redshifts. Points depict EAZY and LePhare redshifts as indicated in the legend. The panel on the right gives the histogram of photometric redshifts using LePhare, which peaks first at the G165 cluster redshift and then again at the redshift of the SN of 1.78 and at the redshift of the DSFG (indicated by the dashed lines).  Bottom panel shows $\vert\delta z \vert/(1+z)$ directly.  
}
\label{fig_photoz}
\end{figure}

Estimates of the photometric redshifts \zph\ were made across the grand mosaic using 
EAZY \citep{Brammer2008,Brammer2021} and LePhare \citep{Arnouts2011}. EAZY SED templates were optimized for the identification of high-redshift galaxies in JWST/NIRCam imaging \citep{Larson2022}. A comparison of the 63 galaxies which have both spectroscopic and photometric redshifts shows good agreement for both approaches. In 95\% of cases, the photometric redshifts are  within 15\% of the spectroscopic redshifts. There are only three outliers present across the two codes (Figure~\ref{fig_photoz}). The one source with a photometric redshift that is too high $\zph \sim6.3$--6.4 compared to the spectroscopic redshift $\zsp= 1.13$ has a secondary redshift solution at $\zph =1.5$ with a nearly equivalent goodness of fit. 
Of the two outliers with photometric redshifts too low, one was found by LePhare at $\zph =0.5$ with $\zsp=3.95$, and one was found by EAZY at $\zph=0.08$ with $\zsp=3.3$. The LePhare outlier is Arc 5.1a, which has a less favored secondary solution at $\zph\sim4.1$, a degeneracy which may be common at this redshift \citep{Frye2023}. 
The EAZY outlier has filter fluxes significantly enhanced by emission lines, LePhare's treatment of which is typically found to perform better \citep{Adams2023,Frye2023}.
 
Following this test of our photometric redshift approach, we extended the photometric redshift estimates to the full multi-band object catalog.   A photometric redshift is considered secure if the object is: (1) in the field of view for all filters, (2)  detected in a minimum of six filters, and (3) spatially resolved from its neighbors. The resulting distribution of photometric redshifts peaks at the cluster redshift and displays minor peaks at the $z\approx1.7$ and $z\approx2.3$  (Figure~\ref{fig_photoz}). The $z=1.7$ bin  corresponds to the redshift of the SN and Arc~2, providing evidence that the SN host galaxy may be one member of a larger galaxy group.  Meanwhile, the peak at $z=2.3$ aligns with the redshift of the lensed DSFG Arc~1, indicating that Arc~1 may be part of another background galaxy group. The physical properties of these two groups are discussed in \S\ref{sec:disc:GGs}.

\subsection{Identification of SN H0pe}\label{sec:phot_SN}
\begin{figure}[h]
\centering\includegraphics[scale=0.19]{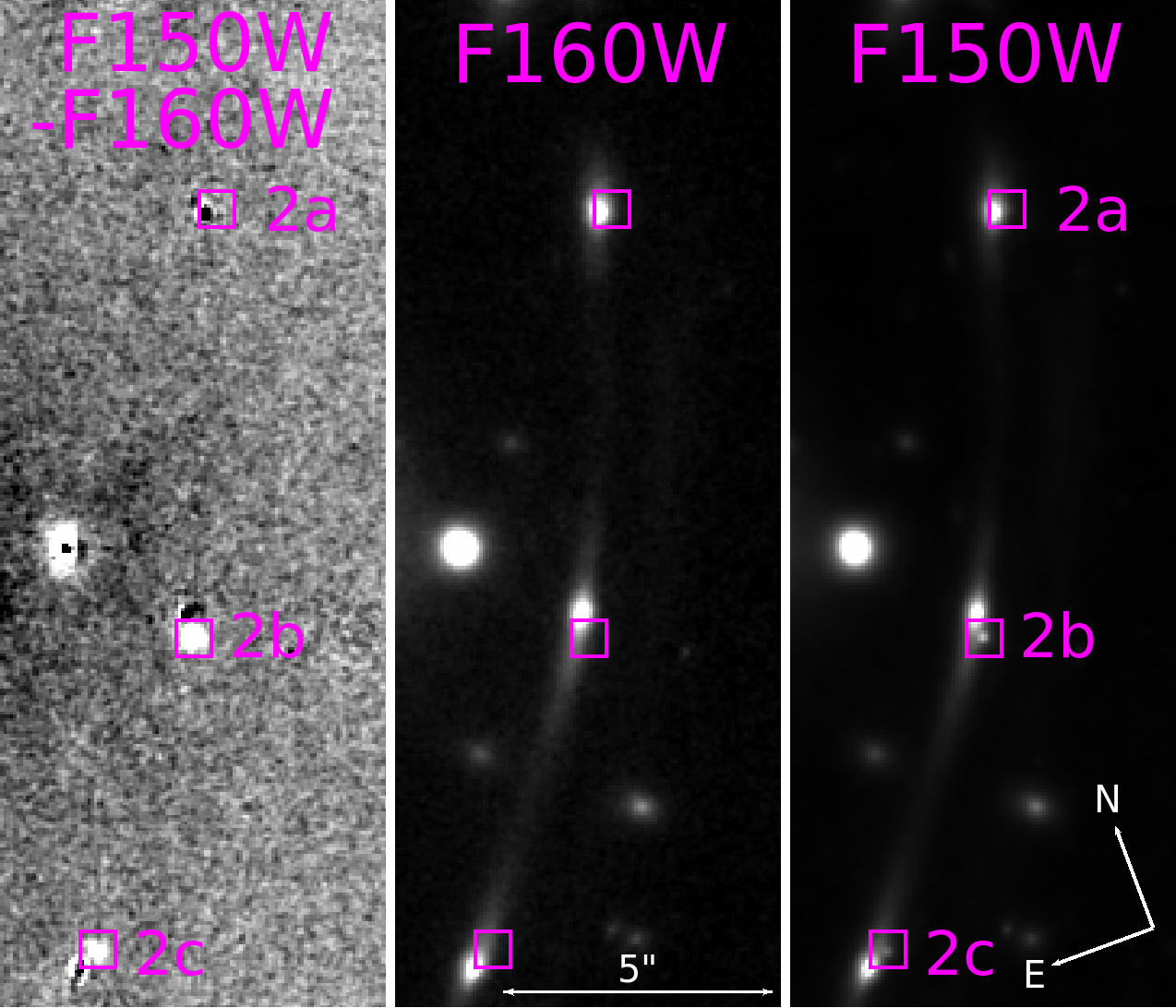}
\caption{Residual image ({\it left panel}) after subtracting HST/WFC3 F160W ({\it middle panel}) from JWST NIRCam F150W ({\it right panel} degraded to match the F160W PSF). This subtraction led to the serendipitous discovery of SN H0pe in all three images of Arc 2. Arc 2b is brightest, followed by 2c and then 2a. 
The lens model predicts that the images arrive in the sequence 2a, 2c, 2b. The HST/WFC3 imaging was obtained on 2016 May 30, and the JWST NIRCam PEARLS imaging was obtained on 2023 March 30.
}
\label{fig_rez}
\end{figure}

As mentioned in \S\ref{sec:outline}, 
three new point-source images were identified in the Epoch~1 imaging. They were close to Arc~2 with a projected source-plane separation estimated from our lens model (\S\ref{sec:LTM}) of 1.5--2~kpc. The separation assumes the point source was at Arc~2's redshift. 
The geometrical arrangement of the point source relative to the arc flipped parity on crossing the critical curve between images Arc~2a and Arc~2b and again between the Arc~2b and Arc~2c images (Figure~\ref{fig_map}), as predicted by lensing theory if the point source is associated with Arc~2. Because the point source follows the same lensing geometry as Arc~2, it is not a spurious source nor a low-redshift interloper and therefore is likely at the same redshift as Arc~2. The triply imaged point source was easily bright enough ($m_{\rm F150W} = 23.91$~AB) to have been detected in the 2016 HST imaging had the source been present then (Figure~\ref{fig_rez}).  The geometry and transient behavior immediately suggested a supernova in the Arc~2 galaxy, and this was confirmed by later observations as detailed below. For this paper, the three images (Figure~\ref{fig_rez}) are referred to as SN~2a, 2b, and 2c.

Figure~\ref{fig_lc} shows the photometry from the Epoch~1 imaging, with the reddest filters (F356W, F444W) omitted due to contamination from the host galaxy. The light curve points, although sparse, were best fit by a Type Ia SN model with $>$90\% probability using the photometric light curve classifier from \citet{Rodney2014}.
This classification predicts a second peak, which for $z=1.78$ will appear at observed wavelengths $\ga$1.8\,\micron. The lens model predicts that the SN~2a image arrived first, and this image is faint in Epoch~1, potentially intercepted already following the the second peak. The model predicts SN~2c should be the second image followed by SN~2b, which apparently was seen at or very near the first peak.  
These initial light-curve fits are only approximations and therefore are not useful for estimating the time delay. At the same time, they were sufficiently compelling to have had Epochs 2 and 3 approved and executed as a disruptive DDT program.

The photometry and spectroscopy in Epochs 2 and 3 confirmed the Type Ia SN designation.  Figure~\ref{fig_stamp} shows images in all three epochs in three representative filters.  As expected based on the lensing predictions, SN~2a was brightest in Epoch~1 in all filters and faded thereafter. SN~2b was the last image to arrive. It was relatively bright in all three epochs because all three are near its light-curve peak.  SN~2c is the intermediate image, which was seen after the first peak even during Epoch~1. Accordingly this image faded in bluer filters but remained bright especially in the LW filters.  The full photometry, the methods for performing the photometry corrected for background galaxy halo light and microlensing effects, and the  photometric time-delay measurements are presented by \citet{Pierel2023H0pe}.  The supernova spectra, the SN type classification, and the spectroscropic time-delay measurement are presented by \citet{Chen2023H0pe}.

\section{Strong Lensing Model} \label{sec:LTM}

\subsection{Inputs for the Lens Model}

\label{sec:constraints}
The light-traces-mass (LTM) model requires as inputs the positions and masses of cluster members, the cluster redshift, and the image systems' identities, positions, and redshifts if known.
We selected cluster members by their spectroscopic redshifts, when available, and augmented this list with sources selected by their near-infrared colors. In particular,  the ``1.6\,$\mu$m bump" is a feature of the stellar populations of $\apg$1\,Gyr galaxies, appearing when massive stars no longer dominate the composite galaxy emission spectrum \citep{Sawicki2002}. For a cluster at $z=0.35$, the 1.6\,$\mu$m bump appears as a positive slope in the $\rm F090W-F150W$ color and a negative slope in the $\rm F277W-F444W$ color. Figure~\ref{fig_bump} shows the color--color selection for G165:
$\rm F090W-F150W\,>\,0.5$ and $\rm F277W\,-F444W\,<\,-0.5$. The galaxies with spectroscopic redshifts reassuringly occupy the expected region of this color--color space. The cluster list is also in agreement with the ``red cluster sequence" method for identifying cluster members \citep{Gladders2000} and enables longer-wavelength selection. The one outlier at $\rm F090W-F150W = 0.35$ and $\rm F277W-F444W = -0.43$ has $z=0.3548$ that places it in the cluster, but this galaxy's photometry was skewed by nebular emission lines.  The main catalog of cluster members contains 161 galaxies enclosed in a region of 3\arcmin\ (924~kpc) on a side centered on (R.A., Decl.) = (11:27:13.9143,+42:28:28.427), depicted by the large green square in Figure~\ref{fig_foot}.

The initial lensing constraints include the 161 cluster members and the 11 known image systems \citep{Frye2019}. New spectroscopic redshifts from  NIRSpec increased the value of image systems 5, 8, and 9 as lensing constraints. A new lens model was then constructed and used to identify new image systems. Table~\ref{tab:long} gives the complete list of image systems. 

\subsection{Light-traces-mass}
In an LTM model, the observed cluster-member light is assumed to trace the underlying dark matter. This semi-parametric approach utilizes few free parameters, making it straightforward to incorporate observational constraints and enable fast computations. This makes the LTM approach especially powerful as a multiple-image finder \citep[][]{Broadhurst2005,Zitrin2009,Zitrin2015}. The LTM method represents each cluster galaxy as a power-law mass surface-density profile with  relative masses set by each galaxy's observed luminosity. The profiles are superposed, and a smoothing kernel is applied to approximate the 
distribution of the dark-matter mass. The amount of dark matter relative to the galaxy light is a free parameter. LTM also includes an external shear component with parameters for its amplitude and position angle. There is additional flexibility to model individual galaxies such as the BCG, where the position angle, ellipticity, and mass-to-light ratio can be left as free parameters to be fit by the model. Image systems that lack spectroscopic redshifts have their redshifts left as free parameters to be fit by the model. 

The LTM model treats all non-luminous mass as a single dark-matter component, i.e., there is no separate component representing gaseous mass.
In fact, G165's gas is not detected in X-rays by ROSAT with an upper limit on the X-ray flux computed from the RASS diffuse map of $1.12\times10^{-4}$\,counts\,s$^{-1}$\,arcmin$^{-2}$ \citep{Frye2019}. Upcoming approved observations with XMM (AO22, \#92030, PI: Frye) should provide the first 2D gas distribution for this cluster for comparison with the stellar distribution. 

The model presented in this study is the first NIRCam-based lens model.
Starting with the base model  of the 11 known image systems \citep{Pascale2022a} and the 161 cluster members, the redshifts for systems 1, 2, 5, 8, and 9 were fixed at their spectroscopic values  (Table~\ref{tab_NIRSpec}), leaving the redshifts of the remaining image systems free to be fit by the model. Additional image systems were then introduced gradually, each time making sure that the fit improved. This process identified 10 new image systems, which we applied as lensing constraints and which are reported in Table~\ref{tab_families}.    

For image systems with multiple components that could be spatially resolved, we used the individual clumps as additional constraints on the lens model. One notable example was SN~H0pe, which provides a separate lensing constraint in addition to the Arc~2 system as a whole. Although Arc~2 is a system of three giant arcs, the galaxy nuclei in each case are compact in F090W, yielding accurate astrometric positions. A second example was Arc~1b/1c, for which Arc~1b separates out into 1.1b and 1.2b, and Arc~1c separates out into 1.1c and 1.2c. 
Additional substructures exist, as indicated in Figure~\ref{fig_map}, but we stopped including these minor constraints when the $\chi$-square did not indicate an improvement in the overall fit. The set of 
substructures used to construct the lens model is in Table~\ref{tab:long}.
 
Figure~\ref{fig_lens} shows the model for which $\chi^2$ is minimized.  This model reproduces the angular positions of input lensed images to an RMS difference of 0\farcs65. While this uncertainty is large compared to the NIRCam pixel size, it is not due to NIRCam WCS errors. Instead it is due to the inability of the lens model to precisely locate all the lensing mass at the right locations in the model. Nevertheless, the  RMS achieved is accurate enough to identify all the plausible counterimages and hence to refine the lens model. 

Based on our lens model, the lensing mass within (projected) 600~kpc of the cluster's luminosity-weighted center is 
$(2.6 \pm 0.30) \times 10^{14}$\,M$_{\odot}$ with the uncertainties computed following the approach of \citep{Pascale2022a}. This value is consistent with the masses estimated by \citet{Pascale2022a} and by \citet{Frye2019} to within the uncertainties.
\begin{figure}[h]
\centering\includegraphics[scale=0.32]{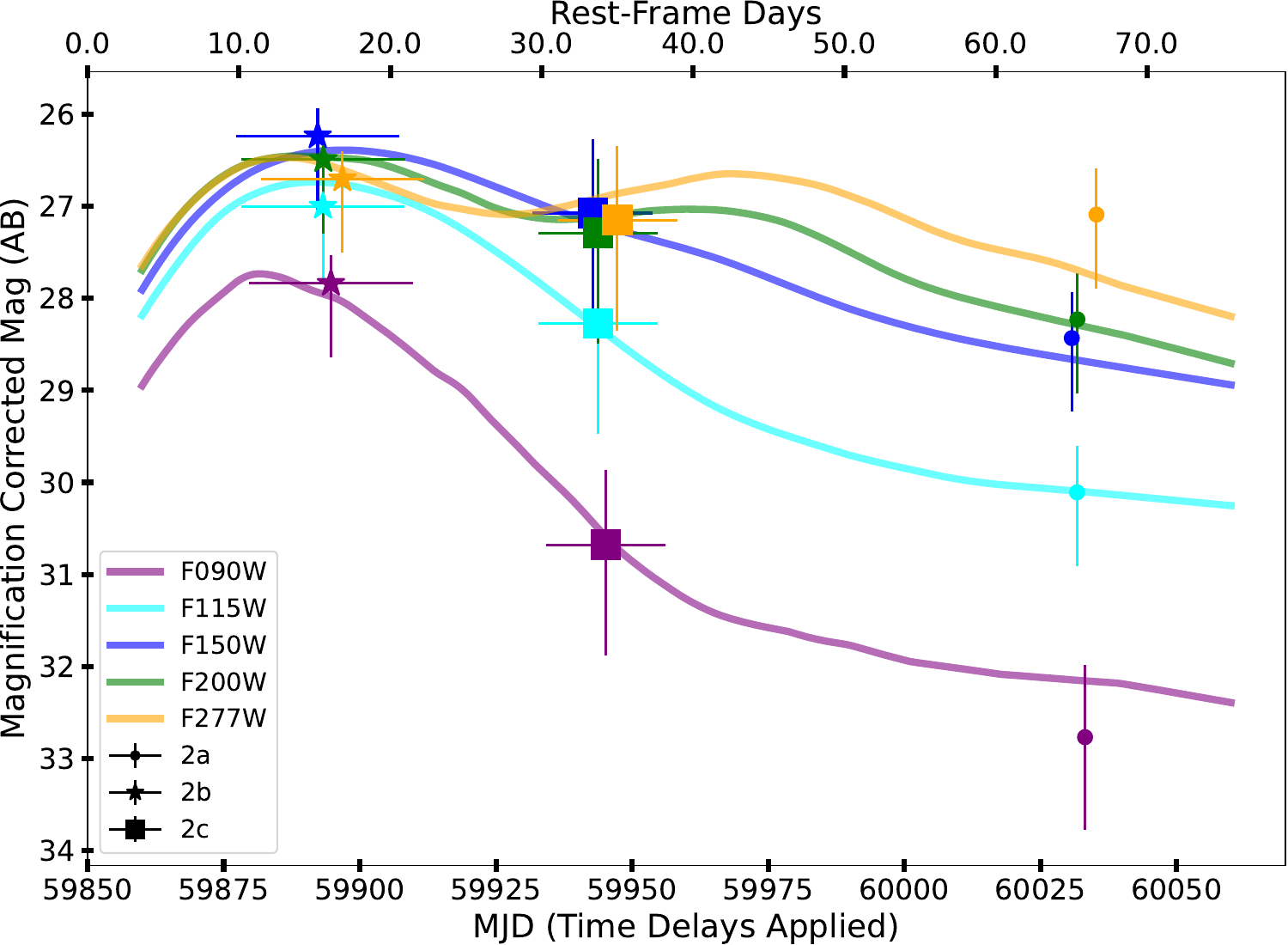}
\caption{Photometry for SN H0pe based on Epoch 1 imaging.  The abscissas are in observed frame (lower) and rest frame (upper) based on time delays derived from the light-curve fitting.
The actual observation epoch was $\rm MJD = 60033$, where the first-arriving image 2a is plotted. The three images gave photometry for the three different times sampled, with SN~2a's arrival followed by SN~2c and then SN~2b, as indicated in the legend. Each point was  corrected for lensing magnification, and images 2b/2c are shifted by the predicted time delay. Points are slightly offset horizontally for clarity (such as for the F277W filter), and lines show best-fit model light curves \citep{Hsiao2007} color-coded by filter as indicated. 
}
\label{fig_lc}
\end{figure}

\begin{figure*}[tb]
\centering\includegraphics[scale=0.44]{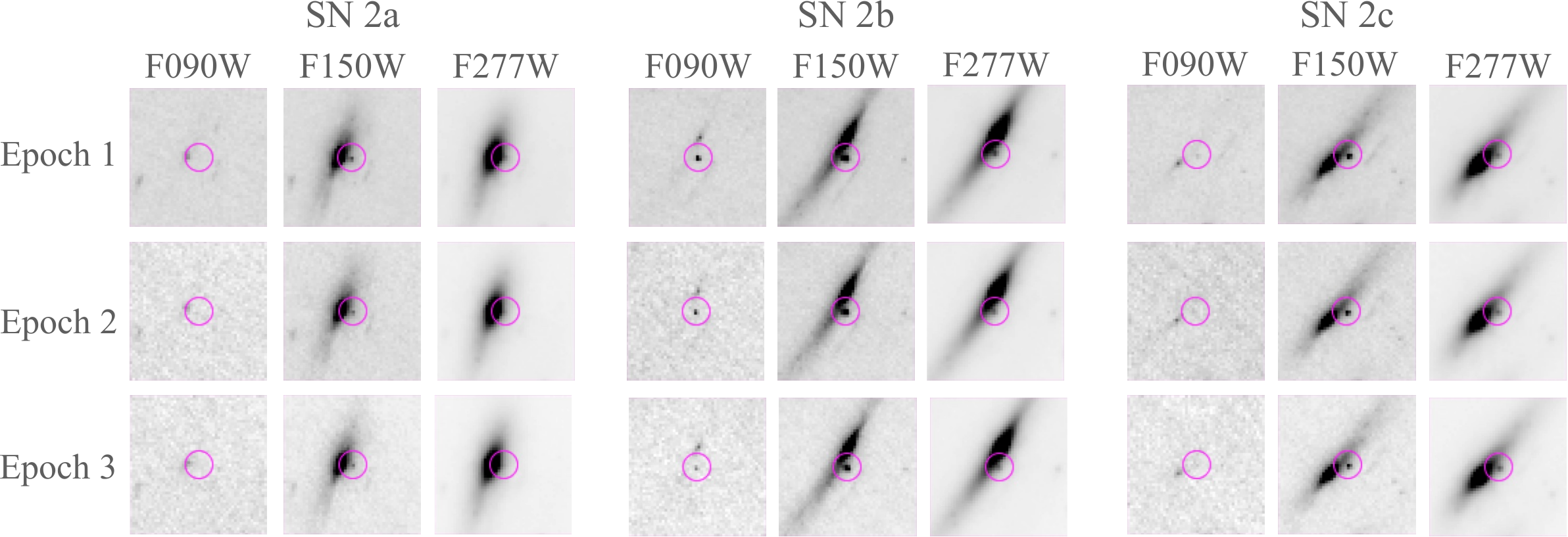}
\caption{Time series of SN H0pe imaging observations. Negative images in the three epochs are shown for each of the three images of the SN\null.  Of the eight NIRCam filters acquired in Epoch 1 and six filters in each of Epochs 2 and 3, only three filters are depicted for clarity. The image stamps are oriented north up, east left and are 3\arcsec\ on a side. The SN appeared first in image 2a, where it was intercepted on the decline after the second IR peak in the light curve. SN 2a and SN 2b were intercepted near the first peak, and Epochs 2 and 3 also trace the second peak detected in the NIRCam LW channels of this Type Ia SN\null.  
}
\label{fig_stamp}
\end{figure*}

\subsection{Mass-Sheet Degeneracy}

Inference of time delays from lens modeling is often limited by the so-called ``mass-sheet degeneracy'' \citep{Falco1985}. The issue is that the modeled lensed image positions, shapes, and magnification ratios are unchanged by adding a sheet of constant surface mass-density to the lens model although the time delays and hence the value derived for $H_0$ can be altered \citep{Schneider2013, Birrer2016, Kochanek2021}. In the case of single-galaxy lenses, where there is typically only a single observed image system, breaking this degeneracy requires kinematic information about the galaxy lens or constraints on the absolute magnification of the images, both of which yield the scaling of the lensing potential. In the case of cluster lenses, the mass-sheet degeneracy can be broken when there is a second image system at a different (spectroscopic) redshift \citep[e.g.,][]{Grillo2020}.

G165 benefits from five image systems with spectroscopic redshifts, mitigating the mass-sheet degeneracy but not completely eliminating it for SN H0pe. The NE component of the cluster, which dominates the lensing of Arc 2, lacks multiple spectroscopically measured image systems at different redshifts, potentially allowing mass-sheet degeneracy effects. Spectroscopy of another image system near Arc~2, such as the caustic-crossing Arc~11, will significantly reduce any effect. Similarly, absolute magnification measurements of SN H0pe via its standard-candle nature may also serve to break this degeneracy. A quantitative analysis of the lens-model-predicted time delays and the effects of the mass-sheet degeneracy is left to a companion paper \citep{Pascale2023}.

\section{Spectroscopic Analysis of the Arc 1 and Arc 2 Galaxy Groups} \label{sec:analysis}

\subsection{Diagnostic Tools} \label{sec:diagnostics}

The NIRCam photometry (\S\ref{sec:NIRCam}) and NIRSpec spectra (\S\ref{sec:obsNIRSpec}) were modeled simultaneously to provide galaxy physical properties and star-forming activity levels. Modeling is based on FAST++   \citep{Schreiber2018,Kriek2009}\footnote{https://github.com/cschreib/fastpp}.  This SED-fitting code incorporated the NIRCam photometry and all of the NIRSpec spectroscopic information (the G140M, G235M, and PRISM data), thereby accounting automatically for any strong emission-line features, which may not be  apparent in the photometry alone. Models are based on a single burst with a range of decay rates $\tau$ and a star formation rate  $\rm SFR(t) \propto \exp(-t/\tau)$, where $t$ is the time since the onset of star formation. The models that provide good fits to the JWST data return information relevant to this study such as the underlying stellar absorption, the dust extinction, and the stellar mass. For galaxies without Balmer emission lines, the dust extinction $A_V$ was extracted from the FAST++ SED fit, and $E(B-V)$ was computed based on the \citet{Calzetti2000} reddening law.
 
Emission lines provide diagnostics on the galaxy classification and activity levels \citep[][and references therein]{Baldwin1981,Curti2020,Curti2023,Li2023,Mingozzi2023}. 
The accessible emission lines are [\ion{O}{2}]$\,\lambda$3727, H$\beta$, [\ion{O}{3}]$\,\lambda\lambda$4959,5007, H$\alpha$, [\ion{N}{2}]$\,\lambda\lambda$6548,6584, and [\ion{S}{2}]$\,\lambda\lambda$6717,6731.
We used {\sc specutils} to measure the line fluxes of H$\alpha$, H$\beta$, and other lines directly from the continuum-subtracted spectrum or via multiple-line fits when H$\alpha$ is blended with 
[\ion{N}{2}]$\,\lambda$6584. These two line fluxes were corrected for underlying stellar absorption based on FAST++ estimates.  Typical values for these equivalent width corrections were $\sim$2\,\AA\ for H$\alpha$ and $<$4\,\AA\ for H$\beta$, both being in the expected ranges \citep{Reddy2018,Sanders2021}. The H$\alpha$ line flux was then corrected for dust extinction based on the measured Balmer decrement  \citep[e.g.,][]{Dominguez2013}. 
For H$\alpha$ and H$\delta$ and for the computation of $D(4000)$ we followed the \citet{Balogh1999} definitions. Otherwise, the line fluxes and equivalent widths typically were measured over a $\sim$1000\,km\,s$^{-1}$ width centered on the line core. Given the grating dispersion of 6.4\,\AA\ pix$^{-1}$, this amounts to $\sim$8 pixels.  Decreasing this velocity width often resulted in a loss of signal near the line tails. This study follows the usual convention that negative equivalent widths indicate emission, and positive values indicate absorption, but exceptions are that H$\alpha$ and [\ion{O}{2}]\,$\lambda$3727 in emission are expressed as positive numbers. 

For galaxies that show strong emission lines, the H$\alpha$ line flux and Balmer decrement give SFR and dust extinction ($A_V$) directly. 
Other emission lines constrain the galaxy activity levels. The emission-line ratios  [\ion{O}{3}]/H$\beta$ ($R_3$), [\ion{N}{2}]/H$\alpha$ ($N_2$), [\ion{S}{2}]/H$\alpha$ ($S_2$), and [\ion{O}{1}]/H$\alpha$ place the galaxy with respect to the \citet{Kewley2006} maximum-starburst line that separates star-forming galaxies (SFGs) from active galactic nuclei (AGNs) \citep[e.g.,][]{Mingozzi2023}. The above line ratios plus 
([\ion{O}{3}]\,5007/H$\beta$)/([\ion{N}{2}]\,6584/H$\alpha$) ($\rm O_3N_2$), ([\ion{S}{2}]\,6716,6730/H$\alpha$)+([\ion{O}{3}]\,5007/H$\beta$) ($\rm RS_{32}$), and ([\ion{O}{3}]\,5007/H$\beta$)/([\ion{S}{2}]\,6716,6730/H$\alpha$) ($\rm O_3S_2$) allow multiple estimates of the gas-phase metallicity.

In some cases, the MSA slit did not fully cover the source galaxy's light, making the measured line fluxes lower limits. 
To compute a correction factor for incomplete coverage of the lensed sources, we performed photometry using the PRISM spectra, which provided uninterrupted coverage of all eight NIRCam bands.  This synthetic photometry was then  compared with the NIRCam photometry, with the uncertainties estimated from the standard deviation of the differences between the eight synthetic photometric values and the measured photometric values. This correction factor was  applied to the values in Table~\ref{tab_GGs} for Arcs 2a and 2c, which had non-negligible slit losses (\S\ref{sec:anal_Arc2}). On applying the corrections for underlying stellar absorption, dust extinction, and incomplete slit coverage of the source, the H$\alpha$ flux was used to derive the SFR using the \citet{Kennicutt1998} relation. The specific star formation rate (sSFR) was then computed by dividing the measured SFR by the stellar mass from  FAST++\null. The stellar mass was computed by the SED fit scaled to the photometry integrated over the entire source and therefore needed no slit-loss correction. The uncertainties on the line fluxes and SFRs stem from the stated uncertainties of the flux values  propagated by bootstrapping with a minimum of 100 realizations with the standard deviation of the distribution yielding the uncertainties. 

\begin{figure}[h]
\centering\includegraphics[scale=0.28]{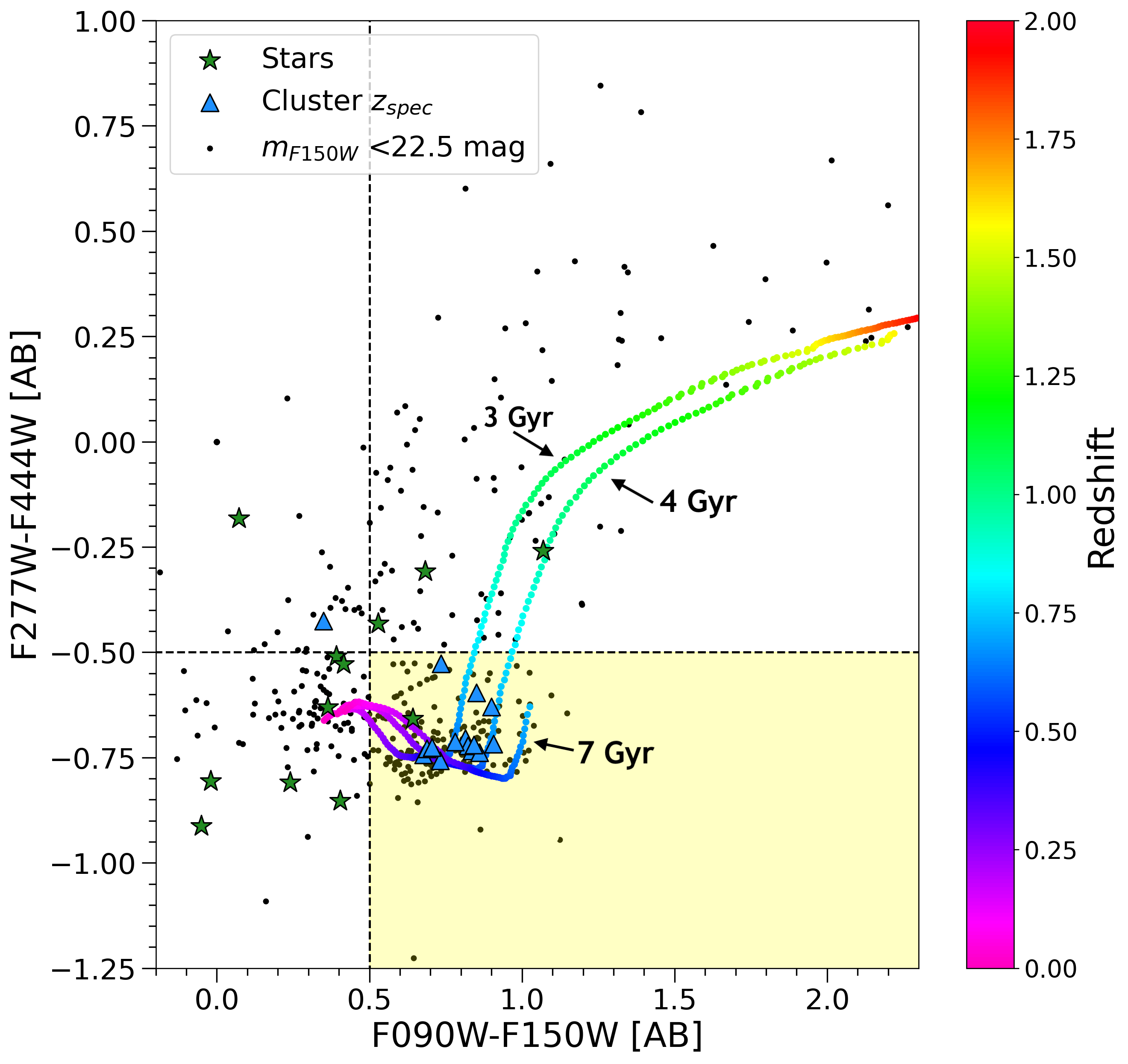}
\caption{NIRCam color--color plot bracketing the rest-frame 1.6\,$\mu$m bump  used to select  G165 members at $z = 0.35$. Small points show galaxy colors measured by NIRCam.  Blue triangles indicate cluster members selected by  spectroscopic redshifts, and green stars indicate Milky Way stars. The yellow-filled region in the lower right shows the color-selection area for cluster members. The colored lines show colors of stellar populations of 3, 4, and 7 Gyrs at $0.1<z<1.5$ with redshift encoded as shown by the color bar. These colors represent a low-dust elliptical galaxy with no AGN\null.
}
\label{fig_bump}
\end{figure}

\subsection{The Arc 1 System: Arcs 1a and NS\_46} 
\label{sec:anal_Arc1}

The Arc~1 system consists of two galaxies, Arc~1a (the DSFG) and NS\_46 both at $\zsp = 2.24$ (\S\ref{sec:obsNIRSpec}, Table~\ref{tab_NIRSpec}). 
Figure~\ref{fig_specDSFG} shows the spectra, 
which both exhibit many emission lines characteristic of starbursting alaxies. 
The H$\alpha$ SFRs corrected for lensing magnification are $\rm SFR= 20^{+40}_{-13}$~\Msol~yr$^{-1}$ and 
80$^{+70}_{-37}$~\Msol~yr$^{-1}$ 
for Arcs~1a and Arc NS\_46, respectively, where our lens model estimates $\mu \approx 2.7$--2.9 for each. 
Both sources are dusty but probably have dust covering-fractions $<$100\% because some rest-frame UV light (observed-frame $g$-band) is detected \citep{Frye2019}. 
The two galaxies are closely separated in radial velocity ($\sim$420 km\,s$^{-1}$ rest-frame) and in projection (5\,kpc), making it tempting to ask whether a galaxy interaction may explain the ongoing star formation.
The measured and estimated internal properties of Arcs 1a and NS\_46 are reported in Table~\ref{tab_GGs}.

\begin{deluxetable*}{cccCCccCc}
\tablecaption{Galaxies in Compact Groups at $z\approx2$}
\tablecolumns{9}
\tablewidth{0pc}
\tablehead{
\colhead{ID} &  \colhead{R.~A.} & \colhead{Decl.} & \colhead{\zsp} & \colhead{$E(B-V)$}   & \colhead{$\mu$} &\colhead{$\log\mu M^{a}$} & $\mu F_{\rm H\alpha}$ & sSFR\\
\colhead{}   & \colhead{(J2000)}      & \colhead{(J2000)}      & \colhead{} & \colhead{}      & \colhead{}                      & \colhead{(M$_{\odot}$)} & \colhead{ ($10^{-17}$\,erg\,s$^{-1}$\,cm$^{-2}$)} & \colhead{(Gyr$^{-1}$)} 
}
\startdata
Arc 1a & 11:27:13.92& +42:28:35.43 &2.2355 \pm 0.0003 & 0.85\pm0.25$^b$&2.9 & 10.4& 19.7^{+37.9}_{-12.7}& $2.3^{+4.5}_{-1.5}$ \\
NS\_46 & 11:27:13.87 & +42:28:35.67 &2.2401 \pm 0.0002 & 0.81\pm0.04 $^b$& 2.7  &9.51& 70.5^{+59.8}_{-32.3}&$67^{+57}_{-31}$\\[0.5ex] 
\hline
Arc~2a   &11:27:15.34&+42:28:41.05& $1.7833\pm0.0010$ &$0.22\pm0.04^c$ & 5.1 & 11.7 & 24.8^{+11.2\,d}_{-8.0}& ${0.076^{+0.040}_{-0.018}}$\\
Arc~2c   &11:27:15.98&+42:28:28.72& 1.7834\pm0.0005 & 0.20\pm0.04$^c$ & 7.2   & 11.7 & 19.4^{+6.9\,d}_{-5.2}  & 
${0.070^{+0.033}_{-0.014}}$ \\ 
NS\_337&11:27:16.28&+42:28:23.59&1.7810 \pm 0.0009 & 0.13\pm0.11$^c$ & 2.5&10.7&\nodata$^e$ & \nodata$^e$\\
NS\_342&11:27:19.46&+42:28:24.79&1.7664 \pm 0.0009 &0.01\pm 0.02$^c$ & 1.6 & 9.18 & 1.9^{+0.3}_{-0.2}&
$2.0^{+0.30}_{-0.20}$\\ 
Arc~8.2c &11:27:15.89&+42:28:28.90&1.7839 \pm 0.0002 &0.24\pm 0.17$^b$ & 9.0 &  8.45  & 2.2^{+5.4}_{-1.5} &$8.5^{+9.8}_{-3.2}$\\
NS\_123 &11:27:04.64&+42:27:15.24 &1.7874 \pm 0.0003 & 0.32\pm0.05$^b$& 1.5 & 9.05& 8.4^{+3.5}_{-2.5}&
$9.6^{+4.3}_{-2.0}$\\ 
Arc~9c   &11:27:16.11&+42:28:27.99&1.7816 \pm 0.0002& 0.31\pm0.05$^b$&5.0&7.43 & 2.8^{+0.8}_{-0.6}&
$260^{+120}_{-55}$\\[1ex] 
\enddata
\tablecomments{Column 1: lensed source name; Column 2: R.~A.; Column 3: Decl.; Column 4: NIRSpec spectroscopic redshift; Column 5: Color excess due to reddening; Column 6: lensing magnification factor estimated from our lens model; Column 7: stellar mass estimated from the FAST++ SED model, uncorrected for lensing magnification $\mu$; Column 8: flux of the H$\alpha$ emission line corrected for underlying stellar absorption and for dust extinction but not for lensing magnification; Arcs 2a/2c have been corrected for incomplete slit coverage;} Column 9: specific star formation rate.
\tablenotetext{a}{Masses are assigned a minimum uncertainty of 0.1~dex to account for systematics \citep{Leja2019a,Leja2019b}.}
\tablenotetext{b}{Measured directly from the Balmer decrement}
\tablenotetext{c}{Derived from FAST++ SED fit}
\tablenotetext{d}{Corrected for incomplete slit coverage of the source}
\tablenotetext{e}{Measurement of H$\alpha$ line flux is not possible for this object.}
  \label{tab_GGs}
\end{deluxetable*}

Our NIRSpec program covered only the DSFG galaxy image pair of Arc~1a and NS\_46 (and not Arcs 1b/1c). The spectrum of each of these arcs yields an H$\alpha$ flux whose sum total implies $\rm SFR(H\alpha) = 50$--210\,M$_{\odot}$\,yr$^{-1}$  (Table~\ref{tab_GGs}). 
If we assume a similar SFR for Arc 1b/c, then the SFR is roughly twice this value.
Additionally, for a DSFG, the dust-obscured SFR may be $\sim$50$-$90\,\% of the total \citep{Whitaker2017}.  In this study we adopt 75\% as a typical value, which would make the total SFR four times the value derived from the unobscured component. Correcting for the absence of NIRSpec data for Arcs 1b/1c and for dust obscuration this translates to a grand total SFR of 400--1700\,M$_{\odot}$\,yr$^{-1}$. Our value for the combined $\rm SFR(H\alpha)$ is roughly consistent with the far-IR, which encompasses all three Arc~1  images and also the nearby galaxy NS\_46 \citep{Harrington2016},
$\rm SFR(FIR) = 400$--800\,M$_{\odot}$~yr$^{-1}$ after demagnification.  Our LTM model gives lensing magnification factors of $\sim$2.9, $\sim$30, and $\sim2.7$ for Arcs 1a, 1b/1c, and NS\_46, respectively.

A condition that makes this comparison challenging is the very different magnifications for the sources. The Arc 1a and NS\_46 images correspond to the total light of the galaxies, whereas Arcs 1b/1c correspond only to a small portion of the galaxy that is multiply imaged. Therefore different regions of the Arc 1 galaxy appear in Arc~1a and Arc~1b/1c, respectively. Moreover, the light from Arc 1a is not easily separable from NS\_46 (Figure~\ref{fig_specDSFG}). Any cross-contamination of the light would alter the SFR tallies. 

Arcs 1a and NS\_46 are classified as star-forming galaxies (SFGs) on account of their high H$\alpha$ equivalent widths, $\gg$10\,\AA\ \citep{Li2023}. Balmer lines are detected in emission all the way up to and including  H$\delta$, and [\ion{O}{2}]\,$\lambda$3727 is strongly detected in emission ($\EW>5$\,\AA), thereby further classifying  these two sources as short starbursting galaxies \citep{Balogh1999}. 
On a longer timescale, the H$\delta$ absorption line that typically appears in galaxies with star formation within the past 800~Myr is undetected in this data set \citep{Goto2007,Weibel2023}.
Taken together with their starbursting-galaxy classifications,  $E(B-V)$ values that approach unity, and high metal enrichment levels, both galaxies in this  pair show some physical properties expected of DSFGs. The bulk of the star formation is recent ($\apll$5\,Myr) and possibly triggered by the interaction of Arc 1a with NS\_46. If this is the case, the galaxy--galaxy interaction may have also instigated any AGN activity.

\subsection{The Arc 2 system: Arcs 2, NS\_337, NS\_342, NS\_123, 8.2c, and 9c} \label{sec:anal_Arc2}

The Arc 2 system consists of Arc 2 (the SN host) and five other galaxies, all at $z=1.78$ (\S\ref{sec:obsNIRSpec}, Table~\ref{tab_NIRSpec}). Figure~\ref{fig_specstack} shows the spectra.
In Arcs~2a and 2c, nearly two dozen emission and absorption line features are identified (Figure~\ref{fig_specstack}). Both sources exhibit H$\alpha$ in emission while H$\beta$ is in absorption, making the Balmer-decrement method of estimating dust extinction unavailable. Therefore for these two galaxy images, the H$\alpha$ line flux was corrected for both underlying stellar absorption and for dust extinction from the FAST++ fit. The H$\alpha$ line flux was further corrected for incomplete slit coverage of the source as described in \S\ref{sec:diagnostics}.  
For Arcs 2a and 2c, we measured correction factors of $1.2\pm0.05$ and $3.9\pm0.26$, respectively. The SFR was then computed and  divided by the stellar mass,  $(5.0 \pm 0.1) \times 10^{11}$~M$_{\odot}$ for both galaxy images. This mass makes Arc~2 the most massive galaxy in the $z=1.78$ group, while its sSFR is the lowest of its cohort.

The most conspicuous difference between the two spectra of this single lensed source is the higher flux density of Arc 2a (Figure~\ref{fig_specstack}).  This is despite the fact that Arc 2c has higher magnification. Most of the difference is incomplete slit coverage of Arc~2c, but there are also minor differences {\it within} each spectrum. As can be seen in the MSA footprint (Figure~\ref{fig_specstack}, right-hand column) different regions of the galaxy are sampled in Arcs 2a and 2c, and in neither one does the slit cover the entire lensed source. In particular, the MSA slit for Arc 2a covers more of the central nuclear region of the arc, which may also contribute to its higher observed flux density.
Some clues may also be drawn from the rather different values for D(4000) of $1.91\pm0.05$ and $1.54\pm0.03$ for Arcs 2a and Arc 2c, respectively. 
It is interesting to ask if this 30\% difference may be a  consequence of the MSA slit sampling a brighter central region that contains a higher fraction of stars in the emerging bulge galaxy component.

The H$\alpha$ emission line indicates that at least some of the star formation is ongoing, and  $\EW(\hbox{[\ion{O}{2}}])>5$\,\AA\ that corroborates this star-forming galaxy classification \citep{Balogh1999}. However, although H$\delta$ is detected in absorption, a feature that is typically associated with galaxies which have undergone star formation within the past 800~Myr, it is weak with $\EW(H{\delta})< 5$\,\AA, possibly indicating that A- and early F-type stars are not yet dominating the spectrum \citep{Goto2007,Weibel2023}. There is a strong sodium~D line, which in combination with the higher $D(4000)$ values indicate the presence of an older stellar population that formed within $\sim$2\,Gyr \citep{Wang2020}. The possibility of a post-starburst (PSB) galaxy subcategory is  ruled out by the presence of H$\alpha$ and [\ion{O}{2}]$\,\lambda$3727 in emission \citep[][and references therein]{Li2023}.   

The sodium D absorption line is stronger in Arc 2 than for any other members of its $z=1.78$ group. The line may consist of both stellar and outflowing-gas components  \citep{Cazzoli2016}.  Another prominent feature that is relatively rare at $z\sim2$ is the detection of all three lines of the calcium triplet (CaT)
associated with cool stellar atmospheres  \citep{Cenarro2003}. Of the possible culprits, supergiants would dominate the luminosity budget, and their presence would be another indicator of a previous star-forming episode in this galaxy's more distant past ($\apg$1\,Gyr). In sum, Arcs 2a and 2c are best fit by a moderately dusty and massive star-forming galaxy that appears to have a complex star-formation history, a result that also matches the findings of \citet{Polletta2022}.

\begin{figure}[h]
\centering\includegraphics[scale=0.12]{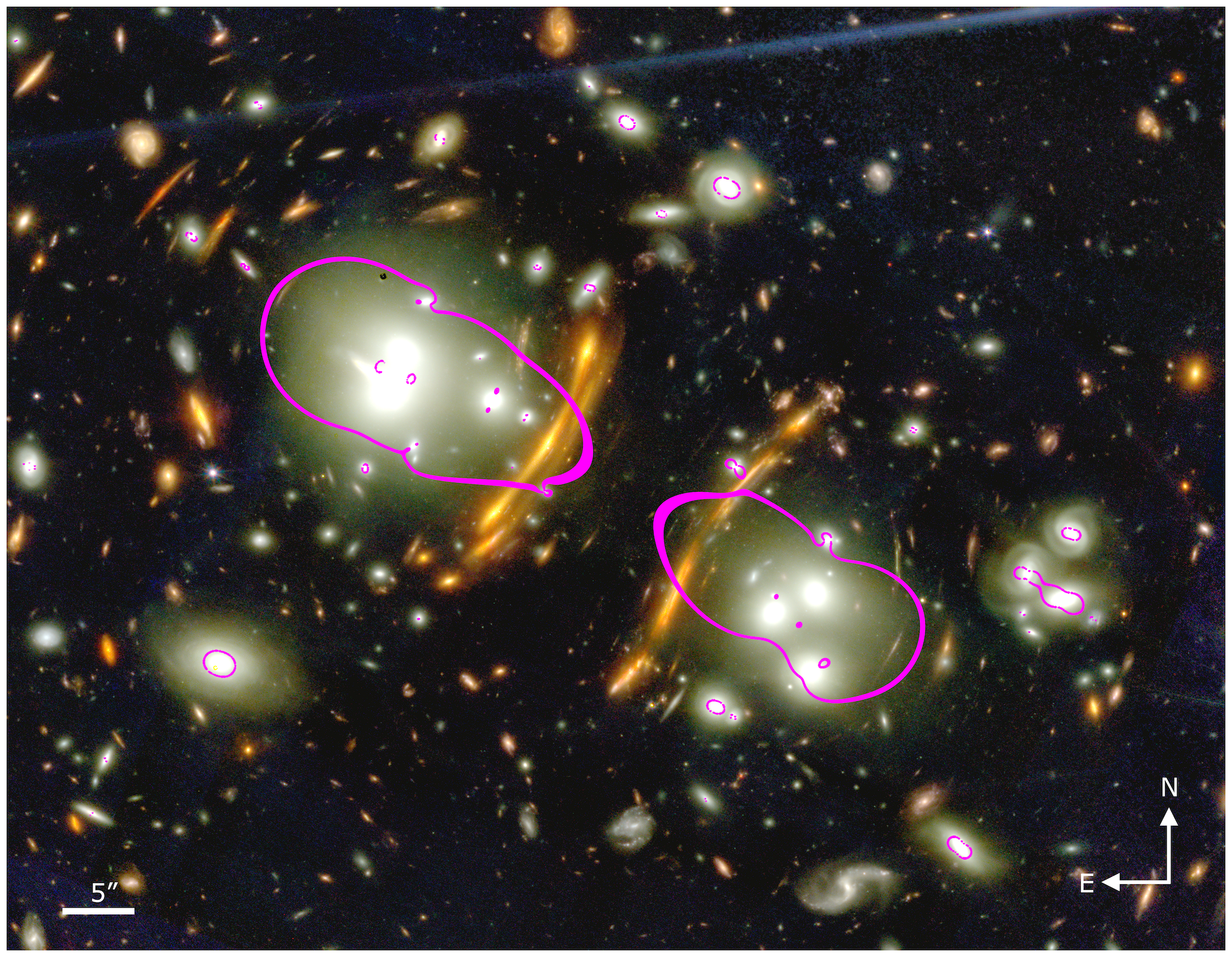}
\caption{Color image using all NIRCam filters showing the central region of the G165 cluster.  Magenta curves depict the $z=2$ tangential critical curve obtained from our LTM model, and white labels show the image scale and orientation. Several giant arcs merge with the critical curve, an attribute that may assist with the detection of transients. The lens model separates out the NE and SW cluster cores at this redshift.  The color rendering is the same as Figure~\ref{fig_specstackimage}. 
}
\label{fig_lens}
\end{figure}

Arc NS\_337 is IR-bright with $m_{F200W}$ = 21.77 AB. The prominent spectral features are detected in absorption except for \ion{He}{1}, which is a common feature detected in nearly all of the $z=2$ spectra in our sample. The computation of the SFR based on the H$\alpha$ emission line is not available for this source. By its $\EW({H\delta}) < 5$\,\AA\ and  $\EW([\hbox{\ion{O}{2}}])<5$\,\AA, this galaxy is classified as passive, although  $D(4000)=1.39 \pm 0.019$ places it near the border with an SFG \citep{Balogh1999}.  The addition of $\EW({H\alpha})<-3$\,\AA\ along with the low $\EW([\hbox{\ion{O}{2}}])$ and $\EW({H\delta})$ also disqualifies it as a PSB galaxy \citep[][and references therein]{Li2023}. 
The absence of H$\alpha$ emission implies at most minimal  ongoing star formation.  The weak H$\delta$  absorption, similar to Arc~2,  indicates that some star formation took place in the past $\sim$800\,Myr. There is a strong sodium D line and a moderately strong  $D(4000)$ break, which indicate the presence of an older stellar population that formed within $\sim$2\,Gyr \citep{Wang2020}. Interestingly, also similar to Arc 2, CaT is detected, possibly uncovering an underlying population of giants or supergiants left over from an older epoch of star formation. The FAST++ fits estimate dust extinction  1~dex lower than Arc~2's and a stellar mass that is also an order of magnitude lower. Overall, Arc NS\_337 is a second relatively massive, quiescent galaxy in this $z=1.78$ group.

Arcs NS\_342, NS\_123, 8c, and 9c are  emission-line galaxies with higher stellar activity levels, making them more akin to the $z=2.24$ group spectra discussed in \S\ref{sec:anal_Arc1}. H$\alpha$ and H$\beta$ are detected in emission for most cases, enabling measurement of the SFR from the H$\alpha$ line fluxes corrected for underlying stellar absorption from a FAST++ fit and for dust extinction by the Balmer decrement method. For Arc NS\_342 where H$\beta$ is not detected in emission, we extracted $A_V$ from the FAST++  fit. These galaxies all have sSFRs  at least an order of magnitude higher than Arc~2's, and their stellar masses are all $\sim$ two orders of magnitude lower.
These four galaxies are all classified as SFGs based on the negative $D(4000)$ values,  $\EW([\hbox{\ion{O}{2}}])> 5$\,\AA\  and $\EW_{H\alpha}> 10$\,\AA\ \citep{Balogh1999,Li2023}. For Arcs 8.2 and 9c, the Balmer  lines are detected in emission all the way through H$\delta$, and $\EW([\hbox{\ion{O}{2}}])>5$\,\AA, further classifying these two sources as short starbursting galaxies \citep{Balogh1999}. In the rest-frame visible, the flux ratios $R_3$, $N_2$, $S_2$, and [\ion{O}{1}]$\,\lambda$6300/H$\alpha$ place Arc NS\_342 and Arc 9c in the star forming (SF) region, Arc 8.2c in the SF/composite region, and Arc NS\_123 in the AGN (subclass Seyfert) region \citep{Mingozzi2023}. Only NS\_123 shows evidence of a predominantly harder source of ionizing flux normally associated with AGNs.

The  rest-frame visible emission lines in Arcs NS\_342, NS\_123, 8c, and 9c enable estimates of their gas-phase oxygen abundances.  The relevant line ratios are defined in \S\ref{sec:anal_Arc1}. (Arc NS\_342 includes only $N_2$ and $S_2$ because H$\beta$ is in absorption even after accounting for underlying stellar absorption and  dust extinction.) Abundances for these four galaxies are  12 + log (O/H) = 8.3--8.6.  These metallicities are higher than the mean metallicity relation derived from  Sloan Digital Sky Survey galaxies \citep[e.g.,][]{Curti2020} for galaxies of the same stellar mass. Together with their high SFRs,  these galaxies may be intercepted during an epoch of rapid build-up of stellar material.  \S\ref{sec:disc:GGs} further discusses the evolutionary state of this compact galaxy group.

\begin{figure}[h]
\centering
\includegraphics[viewport = 360 0 1700 1700,scale=0.14]{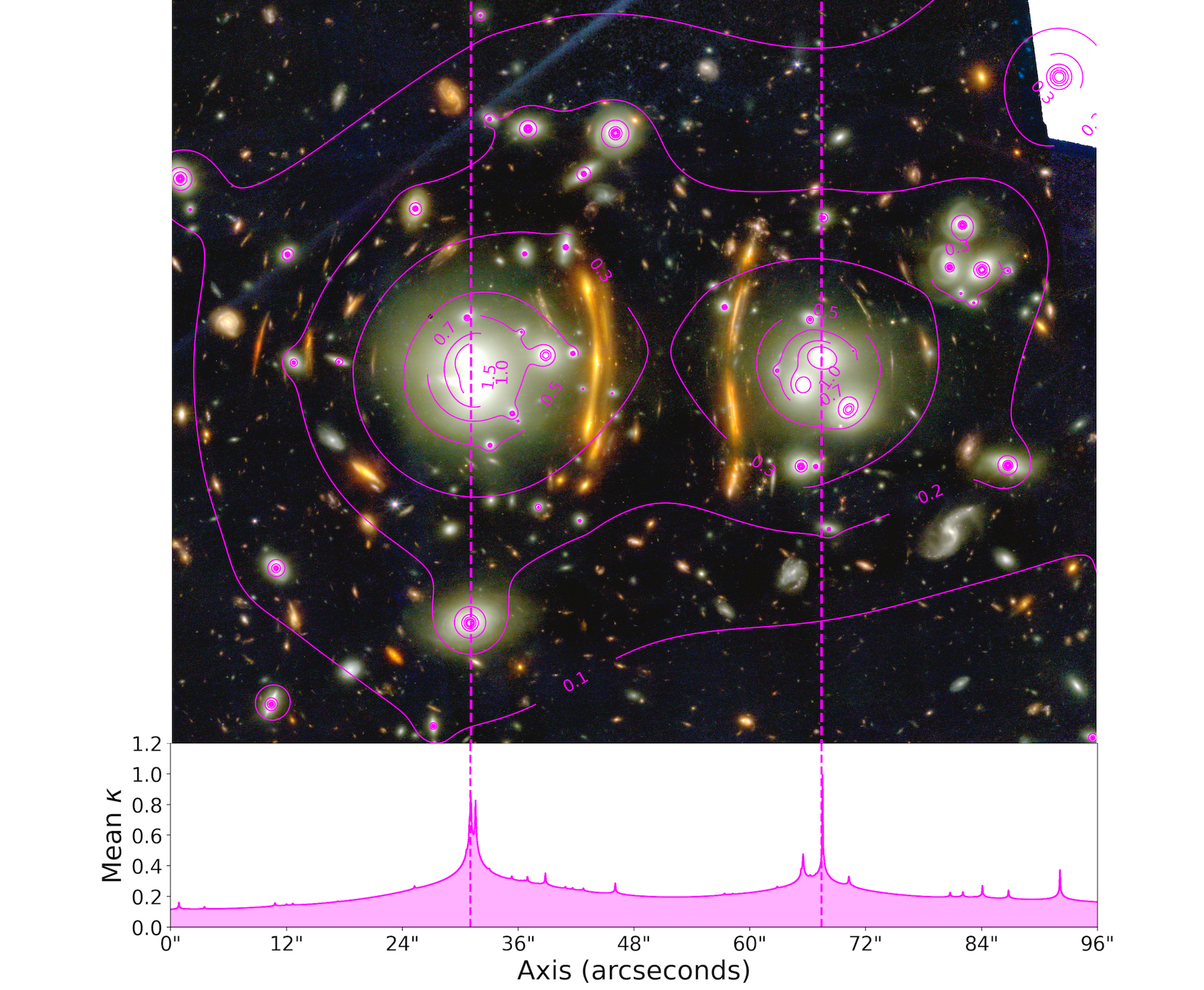}
\caption{G165 mass distribution from our lens model. Pink contours in the upper panel show the surface-mass ($\kappa$) contours scaled to the critical value. The image is oriented as Figure~\ref{fig_map}. The lower panel depicts the 1D mass distribution summed over 78\arcsec\ (=390\,kpc) orthogonal to the line between the mass peaks.  The value for $\kappa$ is summed up over the full field of view depicted in the diagram, which covers an angular range of 96\arcsec\ (=480\,kpc). The NE and SW component mass peaks are marked with long-dashed vertical lines, and there is a modest depression between them. 
}
\label{fig_1D}
\end{figure}

\section{Discussion} \label{sec:disc}

\subsection{G165 Cluster Properties}\label{sec:disc_G165}
In a 1D trace of the surface-mass density (Figure~\ref{fig_1D}), the two peaks correspond to the cluster cores, and each core further separates out into minor peaks that trace individual cluster members. An overall depression, but not a sharp cutoff, appears between the two major peaks. 
The angular separation between the cluster cores is 38\arcsec, 200~kpc at the redshift of the cluster.  Each core contains two radio galaxies, and the two in the NE component are further distinguished by showing extended head--tail morphologies with tails aligned $\sim$parallel to each other  \citep[][and references therein]{Pascale2022a}.  Assuming a radial velocity of 600 km s$^{-1}$ typical of head--tail galaxies \citep{Venkatesan1994} and that the radial velocity  is similar in magnitude to the transverse velocity component, the crossing time would be $\sim$300~Myr. The mean velocity difference of the NE and SW cluster components is not well constrained by current spectroscopic data set \citep{Pascale2022a}, and redshifts of additional cluster members are needed to measure the cluster velocity configuration.

\citet{Pascale2022a} reported a relatively large offset in the centroid position of the luminosity of the spectroscopically-confirmed cluster members which points to a major disturbance of the cluster. The velocity offset of the BCG from the systemic redshift of the cluster is another diagnostic of lack of cluster virialization \citep{Rumbaugh2018}. The BCG's measured redshift of 0.3376 was originally drawn from the SDSS DR 17 archives and is now independently confirmed  (Table~\ref{tab_other_spectra}).
Hence the brightest and largest galaxy in the entire cluster (considering both the NE and SW components) is blueshifted from the cluster's systemic velocity by $\sim$3400 km s$^{-1}$, at face value placing it near the outskirts of the cluster. 

One possible scenario is that the BCG is falling in towards the cluster from behind. This relative newcomer to the cluster would not be detectable by our lens model as a subhalo within the NE component because the redshifts are so similar. Moreover, the BCG is one of the two head--tail radio sources, from which we infer that interactions with the intercluster medium are  imprinting the extended tails in the wake of its motion. This head--tail radio emission is less extended than the other cluster galaxy immediately to the north (\citealt{Pascale2022a}, their Fig.~14). This supports the view that the BCG is primarily falling along the line of sight towards the cluster rather than transversely, although it could also have a less-active nucleus. Another merging cluster, El Gordo, also has the BCG offset by $\sim$2400 km~s$^{-1}$ from the cluster's systemic velocity \citep{Frye2023}.

\subsection{Compact Galaxy Groups \& Associations}\label{sec:disc:GGs}

\begin{figure}[h]
\centering\includegraphics[scale=0.8]{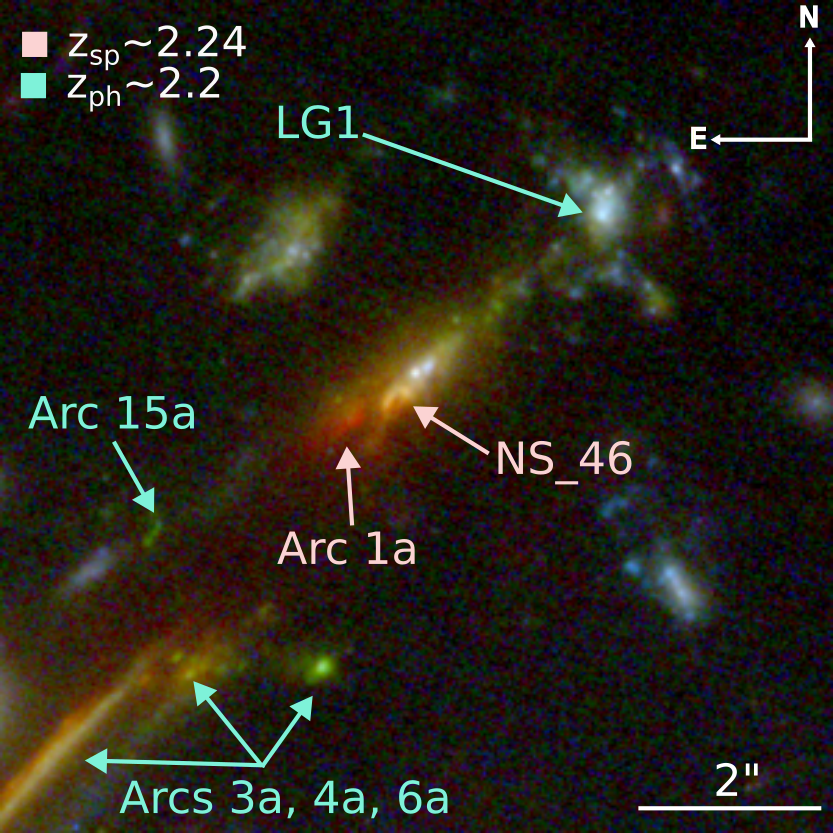}
\caption{The Arc 1a galaxy group at $z\sim2.2$.  Photometric redshift fits identify a compact galaxy overdensity at $z\sim2.2$ that surrounds the spectroscopically confirmed Arcs 1a and NS\_46 consisting of a total of seven objects. These include the multiply imaged Arcs 3, 4, 6, and 15, as well a singly imaged lensed galaxy located only 20~kpc away from Arc 1a and NS\_46. Each galaxy exhibits a clumpy morphology consistent with cosmic-noon starbursts, and if their redshifts are correct, they may be interacting given their $\sim$20~kpc physical separations (after demagnification according to our lens model). Image orientation and scale are labeled, and group members are marked by arrows. The pink arrows indicate the galaxy pair observed in our NIRSpec program\null.}
\label{fig_theclub}
\end{figure}

\begin{figure}[h]
\centering\includegraphics[scale=0.5]{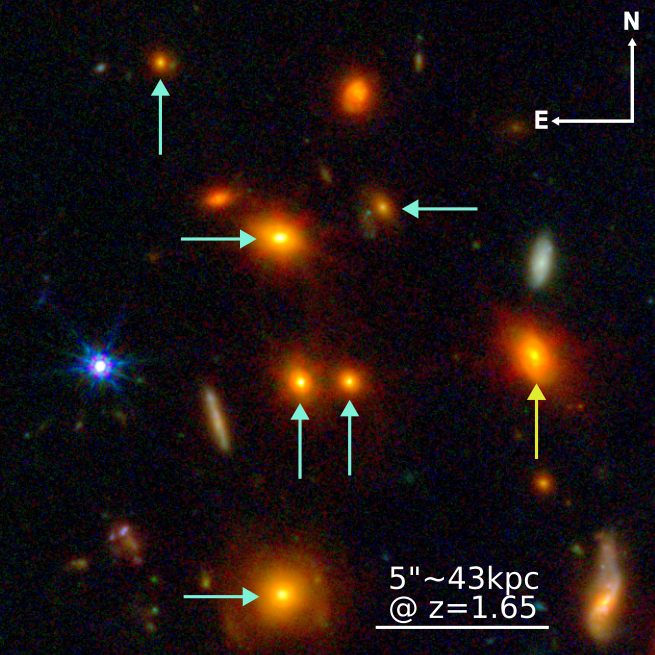}
\caption{
The  $z=1.65$ galaxy group.  Photometric redshift fits identify seven group members. Most candidate galaxy members exhibits similar red colors and elliptical morphologies. These galaxies stand out for their relatively high masses and their ages of $\gtrsim$1~Gyr, as expected of galaxy growth at higher redshifts, followed by gas exhaustion.
Image orientation and scale are labeled, and group members are marked by arrows. The yellow arrow indicates the apparent SFG\null.
}
\label{fig_parallel}
\end{figure}

Three peaks in the background-galaxy redshift distribution (Figure~\ref{fig_photoz}) merit special attention. They correspond to (1) a galaxy association centered in projection on the lensed DSFG Arc 1a at $\zsp=2.24$ (Figure~\ref{fig_theclub}), (2) a galaxy group centered at $\zsp=1.78$ on the SN host Arc 2 (Figure~\ref{fig_specstackimage}), and (3) a galaxy association centered at $\zph=1.65$ in the PEARLS G165 parallel field (Figure~\ref{fig_parallel}). These groups all have  different physical properties, as we discuss below.

{\it The Arc 1 Group:} NIRSpec spectroscopy uncovered a second galaxy proximate to Arc 1a that appears to be the second member of an interacting galaxy pair (Figure~\ref{fig_specDSFG} and \S~\ref{sec:anal_Arc1}). Such interactions are expected to be common at cosmic noon \citep[e.g.,][]{Conselice2006b}. The interaction may explain the SFRs, corrected for lensing magnification, of $\apg$20\,M$_{\odot}$\,yr$^{-1}$ for both (Table~\ref{tab_GGs}). This galaxy pair is flanked by five other galaxies with  $\zph\sim2.2$ (Figure~\ref{fig_theclub}). They include triply-imaged Arc systems 3, 4, 6, and 15 and a single galaxy image that we call LG1. These other galaxies have clumpy morphologies and blue colors consistent with active star formation.  Our lens model predicts that Arcs 1a, 3a, 4a, 6a, and 15a lie within a physical extent in the source plane of 20~kpc.

If the tight configuration of Arc~1-group galaxies is real, then combined with the evidence of starbursting activities in its two central members, it is possible that all seven of these galaxies may be engaged in interactions. The galaxies are expected to grow by star formation perhaps supplied by cold gas streams from the IGM \citep{Dekel2009} and/or by galaxy mergers \citep{Ellison2008,Scudder2012,Ellison2022}.  We may be witnessing the ``preprocessing" of the galaxy members, by which we mean that their stellar masses are being built up prior to virialization and quenching \citep{Rennehan2020,Sengupta2022}. 

{\it The Arc 2 Group:}
By far the dominant peak in the background galaxy redshift distribution is at $z=1.78$ (Figure~\ref{fig_photoz}).
Hundreds of photometrically selected galaxies with this redshift are within the $\sim$1.5 $\times$ 2.5 Mpc field of view of our NIRCam observations. This redshift is so common that it includes about one-quarter of the image systems: Arcs 2, 8, 9, 10, and 16 (Table~\ref{tab_families}). If real,  this  overdense region constitutes a grid of bright $z\sim2$ sources which may explain the abundance of giant arcs and image multiplicities in the field of this lower-mass galaxy cluster.

Of the six spectroscopically confirmed galaxies at $z=1.78$, our lens model estimates that the innermost four Arcs 2, 8, 9, and NS\_337 (labeled in Figure~\ref{fig_specstackimage}) have a physical extent in the source plane of ${\la}33$~kpc with the SN host Arc 2 at the group center. These same four galaxies have a velocity spread of 900 km~s$^{-1}$.  The small spatial and radial velocity extents confirm the presence of a compact galaxy group at $z=1.78$ (Table~\ref{tab_GGs}). 
All but Arc~2 and Arc~NS\_337 have high sSFRs $\apg 1$ Gyr$^{-1}$. Arc~2 has evidence of ongoing star formation, is not a PSB galaxy,  and shows little evidence of star formation in the past $\sim$1Gyr (\S\ref{sec:anal_Arc2}). Moreover, Arc~2 has a stellar mass that is 1--2 orders of magnitude higher than the other spectroscopically confirmed members. The evidence suggests that Arc 2 has already built up the majority of its stellar mass and is now surrounded by star-forming dwarf satellites. This scenario is consistent with the picture of ``downsizing" in hierarchical galaxy formation \citep{Neistein2006,Fontanot2009,Oser2010}.

{\it The $z=1.65$ group:} The third group of galaxies is at $z\sim1.65$ and was uncovered in a photometric redshift search in the NIRCam parallel field (Figure~\ref{fig_parallel}). These seven galaxies exhibit similar red colors, and on visual inspection, all but one have apparent elliptical morphologies. The morphology hints that the red color is a result of older stellar populations rather than dust, a scenario that is supported by our best-fit SED models. If so,  this galaxy association may consist mainly of quiescent galaxies. 
These galaxies have relatively high masses of $10^{10-11}\,{\rm M}_{\odot}$ as estimated by the SED models, stellar ages of $\apg$1\,Gyr, and little to no dust extinction with the exception of the one SFG\null. This is consistent with the scenario of a buildup of stellar material at higher redshifts ($z\apg3$) followed by gas exhaustion at $z\lesssim 1.5$ just prior to an epoch characterized more by hierarchical growth \citep[e.g.,][]{Rennehan2020}.

\subsection{The supernova rate at $z\approx2$}
The total instantaneous (H$\alpha$-based) SFR in the Arc~1 and~2 groups is 
100--500 M$_{\odot}$~yr$^{-1}$.
This rate is corrected for lensing magnification and assumes 75\,\% of the star formation is dust-obscured.
The obscuration correction in \S\ref{sec:anal_Arc1}  is likely to be conservative because these galaxies are especially dusty.
Therefore, the above SFR is likely to be an underestimate of the true value.
In addition, there are other lensed galaxies at the same photometric redshift, such as Arcs 3, 4, 6, and 15, that could be star forming but require spectroscopic confirmation.
The \citet{Young2008} prescription 
implies a lower limit on the rate of core-collapse (cc) SNe $\apg1$--3 SNe~yr$^{-1}$. This value is consistent with results of \citet{Petrushevska2018}, who summed ccSNe over the six Hubble Frontier Fields \citep{Lotz2017}. They found that for a JWST/NIRCam imaging cadence of four 1-hour visits per year, the rate integrated over all six HFFs is expected to be $\sim$0.9 ccSNe and $\sim$0.06 SNe Ia. Regular monitoring of the G165 cluster alone may therefore be well-rewarded with the discovery of additional SNe.

\bigskip
\bigskip

\section{Conclusions and Future Directions} \label{sec:finis}

This overview study presented the full set of JWST observations in the galaxy cluster field of G165. Many of the observations were motivated by a triply-imaged transient identified in the PEARLS Epoch 1 NIRCam imaging.
The three images fit the light curve expected of a SN Type Ia (Figure~\ref{fig_lc}).

We identified 21 image systems in the PEARLS and follow-up data and used them to construct the first NIRCam-based LTM lens model. The new image systems represent nearly a factor of two increase  over previous work and five times as many spectroscopic constraints. Model inputs included 34 cluster members  selected by spectroscopy and 127 selected by the 1.6\,$\mu$m bump. The lens model confirmed the bimodal cluster mass distribution. The model mass within 600~kpc of the cluster center is $(2.6 \pm 0.30) \times 10^{14}$\,M$_{\odot}$, consistent with previous results obtained by the LTM approach. 

NIRSpec spectroscopic redshifts were measured for 30 lensed sources, including all three appearances of the SN, two of the three images of Arc~2, image systems 1, 5, 8, and 9, and other lensed sources. A spectrum of the BCG was also obtained using MMT/Binospec during the {JWST} SN~H0pe observations that confirmed it to be offset from the cluster systemic velocity by $\sim$3400 km~s$^{-1}$, which we take to be evidence of a major cluster disturbance. 
The Arc~1 ($z=2.24$) and Arc~2 ($z=1.78$) image systems appear to be representatives of larger galaxy overdensities at these  redshifts, an assertion backed up by peaks in the photometric redshift distributions.

The two spectroscopic members of the $z=2.23$ Arc~1 system
have relatively high dust levels and high {\it ongoing} star formation expected of starbursting galaxies. The spectra show near-solar gas-phase oxygen abundances. Despite the ongoing star formation, the spectra show little evidence of star formation over $\sim$1~Gyr timescales. In addition to the two spectroscopic group members, photometric redshifts pick out five lensed sources potentially at the group redshift. These sources are characterized by clumpy star formation, and all seven sources are  within a projected extent of 40~kpc. 
We speculate that this compact 
group is being viewed during an epoch of active star formation, a hypothesis that can be tested by obtaining additional and ideally IFU-based spectroscopy to characterize the star formation in the various clumps and knots.

The Arc~2 system at $z=1.78$ has six spectroscopic members.  The triply imaged Arc~2 itself is the SN host galaxy. Arc~2 dominates the group's stellar mass with $M_* = (5.01 \pm 0.10) \times 10^{11}$\,M$_{\odot}$. Although Arc~2 shows evidence for ongoing star formation, it may be petering out compared to rates over $\sim$1~Gyr.  At the same time, Arcs 8.2c, 9c and NS\_123 are emission-line sources which are on average 1--2 orders of magnitude less massive yet have $\sim$2 orders of magnitude higher ongoing sSFRs.
Arcs 2, NS\_337, 8.2c, and 9c are separated in velocity by $\sim$900 km~s$^{-1}$ and are situated within a projected extent of 33\,kpc. 
Arc~2 may be an example of a massive galaxy that completed a star-formation episode $\sim$1\,Gyr ago and is now surrounded by star-forming satellite dwarfs, consistent with a downsizing scenario. 
Spectroscopy is needed to confirm that this photometrically identified galaxy association is a {\it bona fide} galaxy group.

The NIRSpec spectroscopy of SN H0pe will be presented in an upcoming paper that confirms its classification as a Type Ia SN and measures a value for the spectroscopic time delay \citep{Chen2023H0pe}. The NIRCam photometry of the SN measured across all three observing epochs (and hence nine points on the light curve) 
and a photometric measurement of  the time delay will appear in a different paper \citep{Pierel2023H0pe}. And finally, the time delay estimates generated from the lens models, the photometry, the spectroscopy, 
and the weighted combination of all time delay estimates from these models will be used to measure a value for $H_0$ \citep{Pascale2023}. Given the high rate of ongoing star formation across this cluster of $\apg$500 M$_{\odot}$~yr$^{-1}$, regular monitoring of the G165 field may be well-rewarded with the discovery of new SNe and other transients.

\begin{acknowledgments}
This paper is dedicated to PEARLS team member and collaborator Mario Nonino, whose enthusiasm for the science and generosity have been an inspiration for us. 
We thank the two anonymous referees for
suggestions that greatly improved the manuscript.  B.L.F. was funded by NASA JWST DD grant (PID 4446; PI: Frye) from the Space Telescope Science Institute (STScI). B.L.F.~obtained student support through a Faculty Challenge Grant for Increasing Access to Undergraduate Research, and the Arthur L. and Lee G. Herbst Endowment for Innovation and the Science Dean’s Innovation and Education Fund, both obtained at the University of Arizona. R.A.W.~was funded by NASA JWST Interdisciplinary Scientist grants NAG5-12460, NNX14AN10G, and 80GNSSC18K0200 from NASA Goddard Space Flight Center. We thank the JWST Project at NASA GSFC and JWST Program at NASA HQ for their many-decades long dedication to make the JWST mission a success. We especially thank Peter Zeidler, Patricia Royale, Tony Roman, and the JWST scheduling group at STScI for their continued dedicated support to get the JWST observations scheduled. This work is based on observations made with the NASA/ESA/CSA James Webb Space Telescope. The data were obtained from the Mikulski Archive for Space Telescopes (MAST) at the STScI, which is operated by the Association of Universities for Research in Astronomy, Inc., under NASA contract NAS 5-03127 for JWST. These observations are associated with JWST programs 1176 and 4446. This work is also based on observations made with the NASA/ESA {\it Hubble Space Telescope} (HST).  The data were obtained from the {\tt Barbara A.~Mikulski Archive for Space Telescopes (MAST)} at the STScI, which is operated by the Association of Universities for Research in Astronomy (AURA) Inc., under NASA contract NAS 5-26555 for HST.
\end{acknowledgments}

The JWST/NIRCam data used in this paper can be accessed at \url{http://dx.doi.org/10.17909/3zmw-xs33}.  (The proprietary period for the PEARLS NIRCam Epoch~1 data expires on 2024-03-30.) The JWST/NIRSpec data, which are all public, are at \url{http://dx.doi.org/10.17909/hc12-b404}. 

\software{SourceExtractor: \citep{Bertin1996} 
\url{https://www.astromatic.net/software/sextractor/} or
\url{https://sextractor.readthedocs.io/en/latest/}}
{\software{
JWST calibration pipeline version 1.11.2 \citep{pipeline} \url{https://zenodo.org/badge/DOI/10.5281/zenodo.8140011.svg}
}}
\software{
ProFound: \citep{Robotham2018} \url{https://github.com/asgr/ProFound/}}
\software{FAST++  \citet{Kriek2009} \url{https://github.com/cschreib/fastpp}}
\software{SPECUTILS \citet{specutils}}

\facilities{Hubble Space Telescope, James Webb Space Telescope, Mikulski Archive
\url{https://archive.stsci.edu}, MMT/Binospec, MMT/Hectospec, VLA, LOFAR}

\bibliographystyle{aasjournal}
\bibliography{G165_blf}{}

\appendix
\restartappendixnumbering
\renewcommand{\thetable}{A\arabic{table}}
\vspace{-4ex}
Table~\ref{tab:long} lists the G165 arc systems used in the Section~\ref{sec:LTM} lens model.  The image-system designations for numbers 1--11 follow \citet{Frye2019} with the exception of renaming Arc 1a to be the northernmost image and renaming Arc 1b/1c to be the southern images. Ten image systems are new to this study. 
The table columns are: ID, Right Ascension, Declination, observed AB magnitude ({\sc SExtractor mag\_auto}) in the {F200W} filter, spectroscopic redshift (\zsp), photometric redshift (\zph), lens-model predicted redshift ($z_{\rm mod}$), and the arc's discovery citation. Designations with decimal fractions indicate a clump, knot, or other substructure within a larger arc. Positions are J2000 on the GAIA DR3 system. 

\startlongtable
\begin{deluxetable*}{ccccCccc}
\tabletypesize{\footnotesize}
\tablewidth{0pt}
\tablecaption{Strong-Lensing Image Systems} 
\label{tab:long} 
\label{tab_families}
\tablehead{
\colhead{ID}&
\colhead{R.A.}&
\colhead{Decl.}&
\colhead{$m_{\rm F200W,obs}$}&
\colhead{\zsp}&
\colhead{\zph}&
\colhead{$z_{\rm mod}$}&
\colhead{Ref.\tn{a}}
}
\startdata
1a&11:27:13.91&+42:28:35.40&24.90&2.2355 \pm 0.0003\rlap{\tn{b}}&\no&\no& C18, F19 \\ 
1.1b&11:27:14.70&+42:28:23.81&22.18&2.2357\rlap{\tn{cd}}  &\no&\no& C15, H16 \\
1.1c&11:27:14.80&+42:28:21.23&21.83&2.2357\rlap{\tn{cd}}  &\no&\no& C15, H16 \\
1.2b&11:27:14.74&+42:28:23.17&\no&2.2357\rlap{\tn{cd}}  &\no&\no& C15, H16 \\
1.2c&11:27:14.78&+42:28:22.12&\no&2.2357\rlap{\tn{cd}}  &\no&\no& C15, H16 \\
\hline
2a &11:27:15.34&+42:28:41.05&20.30&1.7833 \pm 0.0010\rlap{\tn{be}}&\no&\no &C18, F19 \\
SNa&11:27:15.31&+42:28:41.02&\no\tn{f} &\no\tn{g}&\no& \no&This study\\
2b&11:27:15.60&+42:28:34.19&19.89&$1.783 \pm 0.002\tn{e}$&\no&\no &C18, F19 \\
SNb&11:27:15.60&+42:28:33.73&\no\tn{f} &\no\tn{g}&\no&\no  & This study\\
2c&11:27:15.98&+42:28:28.72&20.26&$1.7834 \pm 0.0005\tn{b} &\no& \no&C18, F19 \\ 
SNc&11:27:15.94&+42:28:28.90 &\no\tn{f}&\no\tn{g}&\no&\no& This study \\
\hline
3.1a&11:27:14.19&+42:28:32.15& \no &\no&\no&2.21& C18, F19 \\
3.1b&11:27:14.24&+42:28:31.46&22.32&\no& 2.2 &\no& C18, F19 \\
3.1c&11:27:14.97&+42:28:17.37&\no  &\no&\no&\no& C18, F19 \\
3.2a&11:27:14.11&+42:28:32.63&\no  &\no&\no&\no& C18, F19 \\
3.2b&11:27:14.36&+42:28:29.62&\no  &\no&\no&\no& C18, F19 \\
\hline
4a&11:27:14.07&+42:28:32.66&22.96&\no& 2.25 &2.173& C18, F19 \\
4b&11:27:14.37&+42:28:29.06&22.77&\no&\no&\no& C18, F19 \\
4c&11:27:14.92&+42:28:17.29&23.62&\no&\no&\no& C18, F19 \\
\hline
5.1a&11:27:13.20&+42:28:25.73&25.36&$3.9530 \pm 0.0004\tn{b}&\no&\no& F19 \\
5.1b&11:27:13.20&+42:28:24.62&24.88&\no&\no&\no& F19 \\
5.1c&11:27:14.27&+42:28:08.83&26.14&\no&\no&\no& F19 \\
\hline
6a&11:27:13.94&+42:28:32.75&24.62&\no& 2.18 &2.215& C18, F19 \\
6b&11:27:14.36&+42:28:27.64&24.32&\no&\no&\no& C18, F19 \\
6c&11:27:14.84&+42:28:16.84&24.89&\no&\no&\no& C18, F19 \\
\hline
7a&11:27:15.24&+42:28:39.46&26.47&\no& 2.45 &1.859& F19 \\
7b&11:27:15.48&+42:28:33.71&24.13&\no&\no&\no& F19 \\
7c&11:27:15.83&+42:28:28.33&24.23&\no&\no&\no& F19 \\
\hline
8.1a&11:27:15.27&+42:28:40.93&24.96\tn{h}  &\no&\no&\no& C18, F19 \\
8.1b&11:27:15.61&+42:28:33.04&22.99\tn{h}  &\no&\no&\no& C18, F19 \\
8.1c&11:27:15.87&+42:28:29.17&24.83\tn{h}  &\no&\no&\no& C18, F19 \\
8.2a&11:27:15.28&+42:28:40.66&24.96\tn{h}  &\no&\no&\no& C18, F19 \\
8.2b&11:27:15.58&+42:28:33.50&22.99\tn{h}  &\no&\no&\no& C18, F19 \\
8.2c&11:27:15.89&+42:28:28.84&24.83\tn{h}  &$1.7839 \pm 0.0002\tn{b}$&&& C18, F19 \\
\hline
9a&11:27:15.43&+42:28:40.57&25.54&\no&\no&\no& C18, F19 \\
9b&11:27:15.57&+42:28:36.02&25.33&\no&\no&\no& C18 F19 \\
9c&11:27:16.11&+42:28:28.01&26.10&$1.7816 \pm 0.0002\tn{b}$&\no&\no& C18, F19 \\
\hline
10a&11:27:15.18&+42:28:38.89&25.93&\no& 1.7 &1.735& F19 \\
10b&11:27:15.46&+42:28:32.86&25.78&\no&\no&\no& F19 \\
10c&11:27:15.74&+42:28:28.51&25.02&\no&\no&\no& F19 \\
\hline
11.1a&11:27:15.77&+42:28:41.92&23.39 &\no&\no&2.46& C18, F19 \\
11.1b&11:27:15.79&+42:28:40.78&23.59 &\no&\no&\no& C18, F19 \\
11.1c&11:27:16.74&+42:28:22.85&25.21\tn{h}&\no&\no&\no& C18, F19 \\
\hline
12a&11:27:14.52&42:28:23.48&27.61&\no&\no&2.581& This study \\ 
12b&11:27:14.73&42:28:17.12&26.27&\no& 2.95 &\no&This study\\ 
\hline
13.1a&11:27:14.04&42:28:37.76&26.86&\no&\no&4.209& This study\\ 
13.2a&11:27:14.07&42:28:37.37&28.52&\no&\no&\no& This study\\
13.1b&11:27:14.75&42:28:27.59&27.89&\no&\no&\no& This study\\
13.2b&11:27:14.71&42:28:28.21&28.68&\no&\no&\no& This study\\
13.1c&11:27:15.14&42:28:18.82&28.89&\no&\no&\no& This study\\
13.2c&11:27:15.16&42:28:18.38&28.80&\no&\no&\no& This study\\
\hline
14.1a&11:27:13.09&42:28:26.83& 24.62\tn{h}&\no& 2.25 &2.342& This study\\
14.1b&11:27:13.20&42:28:18.77&25.16       &\no&\no&\no& This study\\
14.1c&11:27:13.84&42:28:11.32&24.69\tn{h} &\no&\no&\no& This study\\
14.2a&11:27:13.09&42:28:26.94&24.62\tn{h} &\no&\no&\no& This study\\
14.2b&11:27:13.21&42:28:18.51&24.21       &\no&\no&\no& This study\\
14.2c&11:27:13.84&42:28:11.32&24.69\tn{h} &\no&\no&\no& This study\\
\hline
15a&11:27:14.11&42:28:34.10& 26.38    &\no&\no&2.339 & This study\\\
15b&11:27:14.55&42:28:28.27&25.30    &\no& 2.26 &\no& This study\\
15c&11:27:14.96&42:28:19.09&26.75&\no&\no&\no&This study \\
\hline
16a&11:27:13.47&42:28:27.77&24.41 &\no& 1.6 &1.448& This study\\
16b&11:27:13.97&42:28:23.90& 25.20&\no&\no&\no& This study\\
16c&11:27:13.46&42:28:21.70&21.88 &\no&\no&\no&This study \\
16d&11:27:14.22&42:28:13.49&24.51 &\no&\no&\no& This study\\
\hline
17a&11:27:13.23&42:28:28.61&26.92&\no& 6 &6.03& This study\\
17b&11:27:13.22&42:28:23.00&26.89&\no&\no&\no& This study\\
17c&11:27:14.39&42:28:08.12&29.00&\no&\no&\no& This study\\
\hline
18a&11:27:15.35&42:28:37.70&26.27&\no&\no&\edit1{1.83}& This study\\
18b&11:27:15.43&42:28:36.02&26.09&\no&\no&\edit1{1.83}& This study\\
\hline
19a&11:27:14.18&42:28:39.44&27.26&\no&\no&\edit1{4.56}& This study\\
19b&11:27:14.88&42:28:28.69&27.05&\no&\no&\edit1{4.56}& This study\\
19c&11:27:15.30&42:28:20.11&27.10&\no&\no&\edit1{4.56}& This study\\
\tablebreak
20a&11:27:14.44&42:28:35.63&\no&\no&\no&\edit1{3.62}& This study\\
20b&11:27:14.57&42:28:33.55&\no&\no&\no&\edit1{3.62}& This study\\
\hline
21a&11:27:15.09&42:28:35.37&\no&\no&\no&\edit1{1.95}& This study\\
21b&11:27:15.19&42:28:33.31&\no&\no&\no&\edit1{1.95}& This study
\enddata
\tablenotetext{a}{The image system references are: C15 \citep{Canameras2015}, H16 \citep{Harrington2016}, C18 \citep{Canameras2018}, F19 \citep{Frye2019}}
\tablenotetext{b}{Spectroscopic redshift for this arc was measured in this study.}
\tablenotetext{c}{Spectroscopic redshift for this arc was measured by  \citet{Canameras2015} or \citet{Harrington2016}.} 
\tablenotetext{d}{Spectroscopic redshift was measured at a different position along this arc.}
\tablenotetext{e}{Spectroscopic redshift was first measured for this arc by \citet{Polletta2023}.}
\tablenotetext{f}{The photometry is presented elsewhere \citep{Pierel2023H0pe}.}
\tablenotetext{g}{The spectroscopic redshift is assumed to be the same as that of the SN host galaxy Arc~2. The spectroscopy is presented by \citet{Chen2023H0pe}.}
\tablenotetext{h}{Photometry is blended with another image of the same system; the blended photometry is reported.}
\end{deluxetable*}

\end{document}